\documentclass[twocolumn,superscriptaddress,prl,nobibnotes]{revtex4}
\usepackage{mathrsfs, doi, dsfont,mathtools}
\usepackage{amssymb, amsbsy, amsmath, latexsym, mathrsfs,dsfont, array, layout, graphicx,mathrsfs,braket,amsfonts,amsthm, amssymb,graphicx,
 youngtab,color,bm,textcomp,gensymb,csquotes,float,wasysym, siunitx}
\usepackage[caption=false]{subfig}
\usepackage{times}
\usepackage[dvipsnames]{xcolor}

\usepackage{url}

\newcommand{\fig}[1]{Fig.~\ref{#1}}
\newcommand{\tab}[1]{Table~\ref{#1}}

\def\>{\rangle}
\def\<{\langle}

\hypersetup{
    colorlinks=true,  
    linkcolor=black, 
    citecolor=blue, 
    filecolor=blue,
    urlcolor=black
}

\usepackage{bibunits}
\defaultbibliography{OTP_bib}
\defaultbibliographystyle{hieeetr}
\usepackage{hyperref}
\usepackage[all]{hypcap} 
\begin{document}

\begin{bibunit}

\title{Quantum advantage for probabilistic one-time programs}

\author{Marie-Christine Roehsner}
\thanks{These two authors contributed equally}
\affiliation{University of Vienna, Faculty of Physics, Boltzmanngasse 5, 1090 Vienna, Austria}
\author{\normalfont\textsuperscript{\hspace*{-0.1cm}, $\dagger$} Joshua A. Kettlewell}
\thanks{These two authors contributed equally}
\affiliation{Singapore University of Technology and Design, 8 Somapah Road, Singapore 487372}
\affiliation{Centre for Quantum Technologies, National University of Singapore, 3 Science Drive 2, Singapore 117543}

\author{Tiago B. Batalh{\~a}o}
\affiliation{University of Vienna, Faculty of Physics, Boltzmanngasse 5, 1090 Vienna, Austria}
\affiliation{Singapore University of Technology and Design, 8 Somapah Road, Singapore 487372}
\affiliation{Centre for Quantum Technologies, National University of Singapore, 3 Science Drive 2, Singapore 117543}
\affiliation{Centro de Ci\^{e}ncias Naturais e Humanas, Universidade Federal do ABC, Avenida dos Estados 5001, 09210-580, Santo Andr\'{e}, S\~{a}o Paulo, Brazil}

\author{Joseph F. Fitzsimons}
\thanks{Corresponding author}
\affiliation{Singapore University of Technology and Design, 8 Somapah Road, Singapore 487372}
\affiliation{Centre for Quantum Technologies, National University of Singapore, 3 Science Drive 2, Singapore 117543}
\affiliation{Erwin Schr{\"o}dinger Institute, University of Vienna, 1090 Vienna, Austria}

\author{Philip Walther}
\thanks{Corresponding author}
\affiliation{University of Vienna, Faculty of Physics, Boltzmanngasse 5, 1090 Vienna, Austria}
\affiliation{Erwin Schr{\"o}dinger Institute, University of Vienna, 1090 Vienna, Austria}

\begin{abstract}

One-time programs, computer programs which self-destruct after being run only once, are a powerful building block in cryptography and would allow for new forms of secure software distribution. However, ideal one-time programs have been proved to be unachievable using either classical or quantum resources. Here we relax the definition of one-time programs to allow some probability of error in the output and show that quantum mechanics offers security advantages over purely classical resources. We introduce a scheme for encoding probabilistic one-time programs as quantum states with prescribed measurement settings, explore their security, and experimentally demonstrate various one-time programs using measurements on single-photon states. These include classical logic gates, a program to solve Yao's millionaires problem, and a one-time delegation of a digital signature. By combining quantum and classical technology, we demonstrate that quantum techniques can enhance computing capabilities even before full-scale quantum computers are available.

\end{abstract}

\maketitle

With the continuous march of technological advancement, computer processors have become ubiquitous, impacting almost every aspect of our daily lives. Whether being used to compose email or acting as control systems for industrial applications, these devices rely on specially written software to ensure their correct operation. In many cases it would be desirable to prevent a program from being duplicated or to control the number of times a program could be executed, for example to prevent reverse-engineering or to ensure compliance with licensing restrictions. Unfortunately, the very nature of classical information ensures that software can in principle always be copied and rerun, enabling various misuses. 

As a solution to this and other problems the concept of one-time programs was introduced \cite{Goldwasser2008}. One-time programs are a computational paradigm that allows for functions that can be executed one time and one time only. Thus, if a software vendor encodes a function $f$ as a one-time program, a user having only one copy of that program can obtain only one input-output pair $(x,f(x))$ before the program becomes inoperable. In the classical world, this is only possible through the use of one-time hardware or one-time memories \cite{Goldwasser2008}, special-purpose hardware that gets physically destroyed after being used once. However, it is unclear whether such hardware can be realised in an absolutely secure way. An adversary may attack the specific implementation, seeking to circumvent or reverse whatever physical process is used to disable the device after a single use.

Certain features of quantum mechanics, such as the no-cloning theorem \cite{Wooters1982, Dieks1982} and the irreversibility of measurements \cite{vonNeumann55}, suggest that it may enable a solution to this problem. It was recently shown, however, that deterministic one-time programs are impossible even in the quantum regime \cite{Broadbent2013}. As a result, it is believed that neither classical nor quantum information-theoretically secure one-time programs are possible \cite{Goldwasser2008,Broadbent2013,Aaronson09,Mayers97,Lo98,Hayashi06} without further assumptions \cite{Liu2014,Liu2015,erven2014,ng2012,Yao82}.

Here, we demonstrate theoretically and experimentally that quantum mechanics does enable a form of probabilistic one-time program which shows an advantage over any possible classical counterpart. These rely on quantum information processing to execute, but encode entirely classical computation. Such probabilistic one-time programs circumvent existing no-go results by allowing a (bounded) probability of error in the output of the computation. We show that these quantum one-time programs offer a trade-off between accuracy and number of lines of the truth table read, which is not possible in the classical case. Remarkably, the experimental requirements to encode the probabilistic one-time programs we introduce are comparable to those of many quantum key distribution implementations, allowing for technological advances in that field to be harnessed for a new application.
\vspace*{-0.5cm}
\section*{Construction} 
\begin{figure}[t]
\centering
\includegraphics[width=0.5\textwidth]{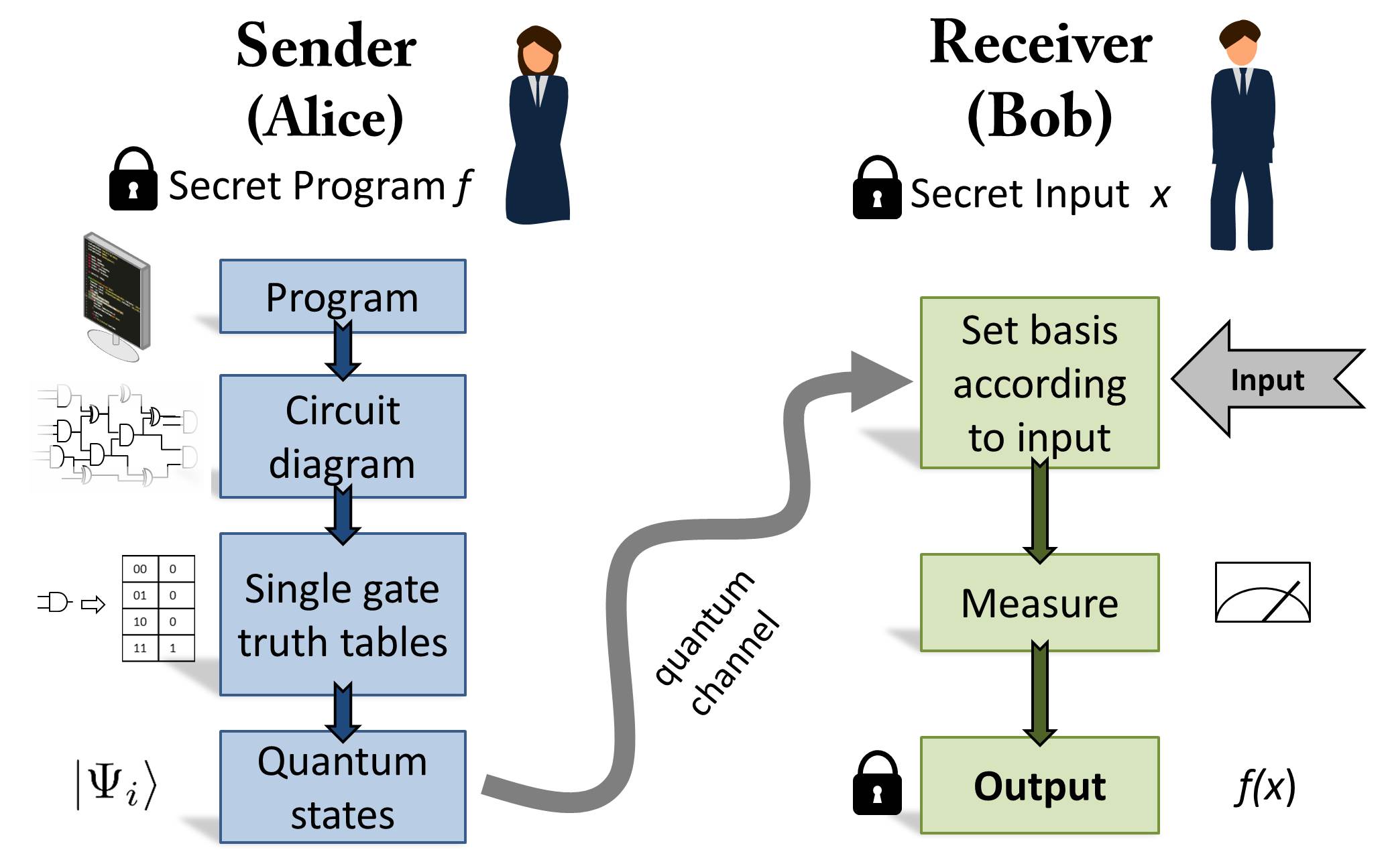}
\caption{Overview of a probabilistic one-time program scheme. Alice possesses a secret program, \textit{f}, and Bob a secret input, \textit{x}. Alice converts $f$ into a logic circuit. Next, Alice encodes the logic gates comprising the circuit as non-orthogonal quantum states. For the particular encoding scheme we realise experimentally, these are always separable states. These states are sent to Bob via a quantum channel. Bob executes the program by sequentially measuring the quantum states corresponding to individual logic gates. The basis for each measurement is determined by Bob's input to that gate and the measurement result represents the output of the gate, up to some bounded probability of error. The encoding can be chosen such that it suffices for Bob to make only single-qubit measurements. Intuitively, the security of the scheme stems from the fact that the measurements corresponding to different inputs for a given gate do not commute, which prevents Bob from evaluating more than one input.}
\label{scheme}
\end{figure}

We consider one-time programs (OTPs) in the context of a two party setting, where Alice is the software provider and Bob is the user. Alice's program is represented by a secret function $f$, which she encodes as a separable state of some number of qubits, which scales linearly in the number of elementary logic gates required to implement $f$, and provides these to Bob. Bob can then evaluate $f$ on some input of his choice $x$ by sequentially measuring each qubit received from Alice. These measurements are a fundamentally irreversible process, which is necessary for Bob to evaluate $f(x)$ while at the same time preventing him from learning $f(x')$ for some input $x' \neq x$. An outline of our approach is presented in \fig{scheme}.

\begin{figure}[t]
\centering
\includegraphics[width=0.5\textwidth]{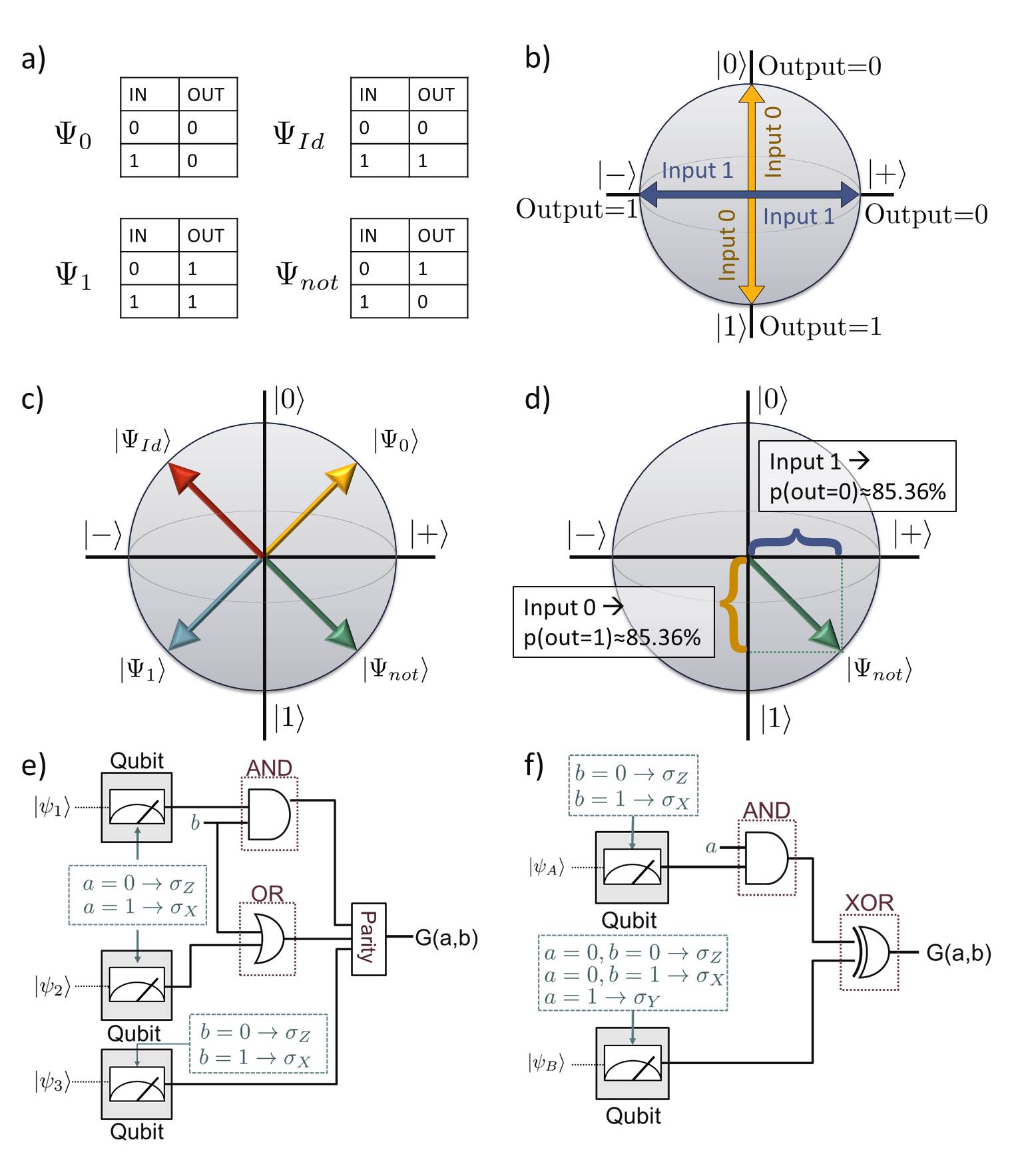}
\caption{Method for constructing probabilistic one-time programs for $\mathcal{G}_1$ and $\mathcal{G}_2$ gates. 
All possible truth tables for gates in $\mathcal{G}_1$ are shown in a) while b) shows the encoding of Bob's inputs: a measurement in $\sigma_Z$ corresponds to an input of $0$ to the gate while a measurement in $\sigma_X$ corresponds to an input of $1$. The output of the gate is given by the measurement outcome. c) shows an overview of the four $\mathcal{G}_1$ states in a Bloch-Sphere representation.
A detailed example is shown in d): when measured in $\sigma_Z$ (i.e. input $0$) the state corresponding to $\ket{\Psi_{not}}$ will be found in the state $\ket{1}$, corresponding to an output of $1$, with a probability of about $85.36\%$.
When measured in $\sigma_X$ (i.e. input 1) the measurement will find $\ket{+}$,  so an outcome of $0$,  with the same probability.
e) and f) show the two equivalent circuits to build an arbitrary $\mathcal{G}_2$ gate.   
e) is based on three $\mathcal{G}_1$ gates while the circuit shown in f) only requires two quantum states per gate, some of which however need to be outside the X-Z-plane.  } 
\label{states}
\end{figure}

In analogy to the compiling of standard classical programs, the logic of $f$ is mapped onto a logic circuit using basic logic synthesis \cite{Lavagno16}. It is necessary that the circuits have a certain standard form, such that the information to be hidden is encoded in the precise choice of logic gates and not on the connections between gates. This is because our approach is to encode the truth table for individual gates as a one-time program in its own right, which we will call {\em gate one-time programs} (gate-OTPs). The interconnection of gates is left public, allowing Bob to propagate information from one gate to the next. Each logic gate is a Boolean function, taking $k$ input bits and returning a single output bit. We will denote the set of $k$-input gates as $\mathcal{G}_k$. For $k\geq 2$, it is possible to implement an arbitrary Boolean function on $n$ input bits with gates chosen only from $\mathcal{G}_k$ together with the fan-out operation \cite{peirce1880} that defines the number of output bits. It is however possible to build up arbitrary $\mathcal{G}_k$ gates from a fixed configuration, with some choice of gates from $\mathcal{G}_1$. Such a construction of an arbitrary $\mathcal{G}_2$ gate  is shown in \fig{states}e.

Probabilistic versions of the four gates comprising $\mathcal{G}_1$ can be encoded using a single qubit, as shown in \fig{states}a-d, such that the measurement operators corresponding to different inputs anti-commute. This is achieved by first fixing the measurement bases corresponding to inputs of $0$ and $1$ respectively (\fig{states}b), and then finding the states to encode each gate such that it maximises the average probability of obtaining the correct outcome across both possible inputs (\fig{states}c). The measurement bases are chosen to be unbiased and correspond to anti-commuting observables, $\sigma_Z$ for input $0$ and $\sigma_X$ for input $1$, to ensure that in learning about the value of one observable Bob must forego information on the other. Once the measurement bases are fixed, the states can be found which yield the correct output with a maximal probability of $\frac{1}{2} + \frac{1}{2\sqrt{2}}$ or approximately 85.36\%. This encoding relates to conjucate encoding introduced by Wiesner \cite{Wiesner1983} and is equivalent to the quantum random-access codes considered in \cite{Nayak}, which were motivated by ideas of compression rather than security. However, the concepts of one-time programs and random access codes diverge when we consider hiding gates from $\mathcal{G}_k$ for $k>1$ later on.

With a method for implementing $\mathcal{G}_1$ now in place, we can proceed to construct a universal set of gates, for example $\mathcal{G}_2$, while preventing Bob from learning the full truth table. As alluded to previously, one way to achieve this is to insert hidden $\mathcal{G}_1$ gates into a fixed circuit, as shown in \fig{states}. The exact choices required for each of the hidden gates to achieve a specific $\mathcal{G}_2$ gate is described in the Supplementary Information. The overall success probability for gates constructed in this way is $75\%$. However, such an approach yields a rather complicated construction for gates in $\mathcal{G}_k$ for $k>2$ and introduces complications in the security analysis. A more appealing approach is to directly implement probabilistic one-time programs for gates in $\mathcal{G}_k$. This can be done by generalising the construction used in the $k=1$ case. Specifically, each possible input is assigned a unique observable from a set of anti-commuting multi-qubit Pauli operators $\{\sigma_i\}$, where a $+1$ measurement outcome is taken to correspond to a gate output of $0$ and a $-1$ outcome is taken to correspond to an output of $1$. As before, the states encoding each gate $G$ are chosen to maximise the average probability that the outcome of measuring the observable corresponding to input $x$ results in output $G(x)$. Unlike the case for $\mathcal{G}_1$, there is an entire subspace of states satisfying this constraint for a given $G$. Our approach is to encode $G$ as the maximum entropy state maximising success probability,  
\begin{equation}
\rho_G = \frac{1}{\text{tr}(\mathbb{I})} \left( \mathbb{I} + \frac{1}{\sqrt{2^k}}\sum _{i=0} ^{2^{k}-1} \left(-1\right)^{G(i)} \sigma_i \right) \; .
\label{eq:quantumgatestate}
\end{equation}
This coincides with the definition a particular type of random access code, known as a parity oblivious random access code, explored in \cite{chailloux2016optimal} for other purposes. The success probability for any input $i$ is then given by $\frac{1}{2}\left(1+(-1)^{G(i)}\text{tr}(\rho_G \sigma_i)\right)$ which simplifies to $\frac{1}{2}\left(1+2^{-\frac{k}{2}}\right)$. Remarkably, this results in each $\rho_G$ being the maximally mixed state of a $2^{k-1}$-dimensional subspace, so that the von Neumann entropy is $k-1$.

The implementation of $\mathcal{G}_k$ gates requires $2^k$ anti-commuting operators and $2^{k-1}$ qubits. However, there is an alternative implementation that uses $2^k-1$ qubits whose Pauli operators are restricted to being tensor products of the identity, $\sigma_X$ and $\sigma_Z$.
While there is no fundamental reason to require such a restriction, it can reduce the hardware requirements necessary to implement the scheme, as seen in the experimental section. These encodings form the basis for the experimental implementations with elliptically and linearly polarised photons respectively.

\section*{Explicit gate construction} 

Here we show the explicit form of single-photon states that can be combined to encode all  $\mathcal{G}_1$ and $\mathcal{G}_2$ gates as shown in the Supplementary Information.

\subsubsection*{Gates with 1 bit of input}

The simplest case of program is one that accepts one bit of input and returns one bit of output. 
The truth tables for all such $\mathcal{G}_1$ gates (shown in \fig{states}c) may be easily encoded as:
\begin{subequations}
\begin{align}
\ket{\Psi_0} = \frac{1}{\sqrt{2+\sqrt{2}}} \left( \ket{0} + \ket{+} \right) \; ,\label{linear1}\\
\ket{\Psi_1} = \frac{1}{\sqrt{2+\sqrt{2}}} \left( \ket{1} - \ket{-} \right) \; ,\label{linear2} \\
\ket{\Psi_{{Id}}} = \frac{1}{\sqrt{2+\sqrt{2}}} \left( \ket{0} + \ket{-} \right) \; , \label{linear3} \\
\ket{\Psi_{{not}}} = \frac{1}{\sqrt{2+\sqrt{2}}} \left( \ket{1} + \ket{+} \right) \; ,
\label{linear4}
\end{align}
\end{subequations}
where $\ket{\pm} = \frac{1}{\sqrt{2}}\left( \ket{0} \pm \ket{1} \right)$.

\subsubsection*{Gates with 2 bits of input}

All $\mathcal{G}_2$ gates can be encoded using either a combination of three states from Equations \ref{linear1}-\ref{linear4} (which corresponds to the linear scheme) or a combination of two states (elliptical scheme), in which case the above mentioned states need to be combined with additional states from the following list:
\begin{subequations}
\begin{align}
\left|\Psi_{0}^{e}\right\rangle & =\left(+\frac{1}{2}-\frac{1}{\sqrt{2}}i\right)\left|0\right\rangle +\frac{1}{2}\left|1\right\rangle \label{elliptical1} \; ,\\
\left|\Psi_{1}^{e}\right\rangle & =\left(-\frac{1}{2}-\frac{1}{\sqrt{2}}i\right)\left|0\right\rangle +\frac{1}{2}\left|1\right\rangle \; ,\\
\left|\Psi_{2}^{e}\right\rangle & =\frac{1}{2}\left|0\right\rangle +\left(+\frac{1}{2}+\frac{1}{\sqrt{2}}i\right)\left|1\right\rangle \; ,\\
\left|\Psi_{3}^{e}\right\rangle & =\frac{1}{2}\left|0\right\rangle +\left(-\frac{1}{2}+\frac{1}{\sqrt{2}}i\right)\left|1\right\rangle \; ,\\
\left|\Psi_{4}^{e}\right\rangle & =\left(+\frac{1}{2}+\frac{1}{\sqrt{2}}i\right)\left|0\right\rangle +\frac{1}{2}\left|1\right\rangle\; ,\\
\left|\Psi_{5}^{e}\right\rangle & =\left(-\frac{1}{2}+\frac{1}{\sqrt{2}}i\right)\left|0\right\rangle +\frac{1}{2}\left|1\right\rangle \; ,\\
\left|\Psi_{6}^{e}\right\rangle & =\frac{1}{2}\left|0\right\rangle +\left(+\frac{1}{2}-\frac{1}{\sqrt{2}}i\right)\left|1\right\rangle \; ,\\
\left|\Psi_{7}^{e}\right\rangle & =\frac{1}{2}\left|0\right\rangle +\left(-\frac{1}{2}-\frac{1}{\sqrt{2}}i\right)\left|1\right\rangle\; .\label{elliptical8}
\end{align}
\end{subequations}

The encoding of specific gates is done according to tables shown in the Supplementary Information. In the linear and elliptical scheme, the gate-encoding state is a tensor product state of three or two photons, respectively. In the linear scheme, each of the three photons are in a state given in Equations \ref{linear1}-\ref{linear4}. As there are 64 combinations and only 16 gates, each gate can be encoded in four different ways (represented by orthogonal state vectors), and a random choice is made each time the gate must be encoded. In the elliptical scheme, the first photon is in a state given in Equations \ref{linear1}-\ref{linear4}, while the second photon is in a state given in Equations \ref{elliptical1}-\ref{elliptical8}. As there are 32 combinations and only 16 gates, each gate can be encoded in two different ways, and again a random choice is made each time the gate must be encoded. The random choice between orthogonal state vectors is made by the sender and it is irrelevant from the point of view of the receiver. Thus, the state as seen by the receiver is effectively the mixed state given in Equation \ref{eq:quantumgatestate}.

\section*{Experimental implementation}

\begin{figure}[t]
\centering
\includegraphics[width=0.5\textwidth]{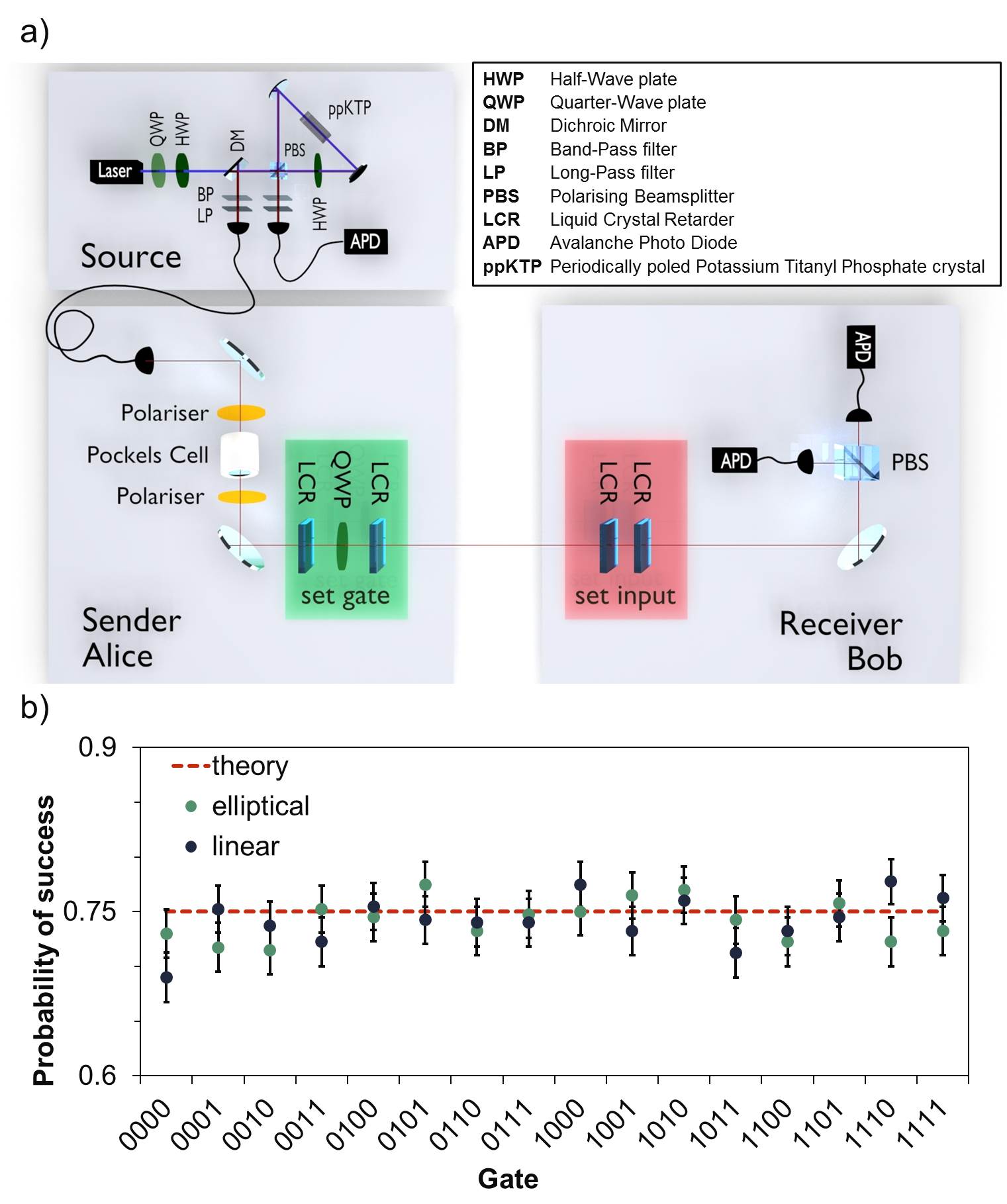}\caption{Experimental one-time program implementation. 
a) shows our setup:  Marked in green are the liquid crystal retarders (LCR) and quarter-wave plate (QWP) used to manipulate the polarisation state of single photons which corresponds to setting the individual gates of the encoded program. 
The LCRs marked in red are used by Bob to set the measurement basis according to the gate inputs. On Alice's side an active switch is implemented based on a Pockels cell placed between two crossed polarisers acting as a fast switching half-wave plate. Alice produces single photons with a source in a Sagnac configuration \cite{Kim2006,Fedrizzi2007, Procopio2015}. In b) we show the average success probability per gate for all $\mathcal{G}_2$ gates.
Blue dots represent the results for the linear scheme, green dots the results for the elliptical scheme and the theoretically predicted value is shown by the red line. 
The gates are labelled by the last column of their truth table, e.g. $(00{\rightarrow} 0, 01{\rightarrow}1, 10{\rightarrow}0, 11{\rightarrow}0)$ corresponds to $0100$. }
\label{Setup}
\end{figure}

To demonstrate the viability of the presented scheme we show a proof-of-principle implementation based on polarisation encoded photonic qubits (\fig{Setup}a).
 
We realized two equivalent schemes: we refer to the first one as the \textit{linear scheme} because it can be implemented using only linearly polarised photons. This version requires fewer technological resources: Alice and Bob each need just one liquid crystal retarder (LCR). These LCRs rotate the polarization of each photon by an angle depending on the applied voltage and are therefore used to actively switch from one polarisation setting (corresponding to a gate or a measurement basis) to the next. However, in this encoding three photons per $\mathcal{G}_2$ gate are required. Our \textit{elliptical scheme} uses elliptically polarised states and requires two LCRs per party. The advantage of this scheme is that it only requires two photons per $\mathcal{G}_2$ gate, reducing the length of the program by a third. 

For both versions we tested all 16 gates comprising $\mathcal{G}_2$ for all four possible inputs ($00$, $01$, $10$, $11$). The average success probability of each gate is shown in \fig{Setup}b, and the results are in good agreement with the expected value of $0.75$. We characterized all single-photon states using quantum state tomography \cite{Munro} where a fidelity,  $F\geq 0.991\pm 0.008$ could be achieved for all states (see \tab{Fidelities} for details). 
 
 \section*{Experimental setup}

Our single-photon source is based on spontaneous parametric down conversion (SPDC) using a Sagnac loop \cite{Kim2006,Fedrizzi2007, Procopio2015}. The pump beam is generated by a \SI{4.5}{\milli\watt} diode laser at a central wavelength of \SI{394.5}{\nano\metre}, followed by a half- and a quarter-wave plate to adjust the polarisation. It was focused on a \SI{20}{\milli\metre} long, type-II colinear periodically poled Potassium Titanyl Phosphate crystal placed inside the loop, which emitted photon pairs at \SI{789}{\nano\metre} in a separable state $\ket{H}\ket{V}$, where H and V denote horizontal and vertical polarization respectively. The down-converted photons were reflected by a dichroic mirror while the pump beam was transmitted. Additionally long-pass and band-pass filters were used to block the pump beam and to select the desired wavelength for the photon pairs. The down-converted photons were then coupled into single-mode fibres and one was directly sent to a detector to herald the second photon. The source was configured in a way that we observed a typical two-photon coincidence-rate of \SI{2}{\kilo\hertz} with an open switch and the ratio of multi-pair events to single-pair events was \SI{<0.07}{\percent}. 
The possibility of multi-pair emissions is a property of every SPDC process which in our case could lead to the transmission of more than one photon at once through the switch and therefore cause unwanted information leaking to the client. Should a future application require even lower (or vanishing) multi-pair emission this could be implemented using alternative single photon sources \cite{Eisaman2011,Schweickert2017,somaschi2016,senellart2017}.

Furthermore we implemented an active switch based on a KD*P (potassium dideuterium phosphate) Pockels cell with a half-wave voltage of \SI{6.3}{\kilo\volt} and two crossed polarisers. The electronic signal from the avalanche photo diode detector (APD) in the heralding path was sent to a splitterbox which could produce an \textit{on} and \textit{off} signal for the driver of the Pockels cell. The pulses were separated by \SI{46}{\nano\second} which corresponds to the opening time of the switch. During this time voltage is applied to the Pockels cell, causing it to act as a HWP. 

These pulses are gated to ensure photons are not transmitted while the LCRs are changing. Once the LCRs are ready to set a state in the program, a gating signal is sent to the splitterbox. Only then will the next heralding signal cause an on/off pulse to be sent to the Pockels cell. All following herald signals will be blocked until the splitterbox receives the next gate signal.

The splitterbox itself causes a delay of the electric signal of \SI{22}{\nano\second} while the total electronic delay of splitterbox and control electronics is \SI{80}{\nano\second}. The Pockels cell has a rise-time of \SI{8}{\nano\second}. To allow for the switch to be opened before the signal photon reaches the Pockels cell in spite of all electronic delays the signal photon is delayed in a \SI{29}{\metre} single mode fibre. 
All necessary polarisation states were set using a combination of two LCRs and a QWP at \SI{0}{\degree}. The maximum time to switch between two states in our scheme was \SI{60}{\milli\second}. This was therefore the time allowed for every switching process (so as not to leak information about the prepared state because of a shorter switching time). 
To measure the states in the bases dictated by the inputs to the gates a second set of two LCRs was used followed by a PBS and two APDs to measure the photons. Typically \SI{4}{\percent} of the times the switch opened a photon was also detected at Bob's side. This was due to losses in the setup as well as the limited detection efficiency of the APDs. Together with the LCR switching time of \SI{60}{\milli\second}  this lead to an of average gate time of \SI{1.4}{\second} per photon. 

\section*{Demonstrated programs}

To demonstrate the applicability of our scheme we have experimentally implemented two different classes of one-time programs. 

\begin{figure}[t]
\centering
\includegraphics[width=0.48\textwidth]{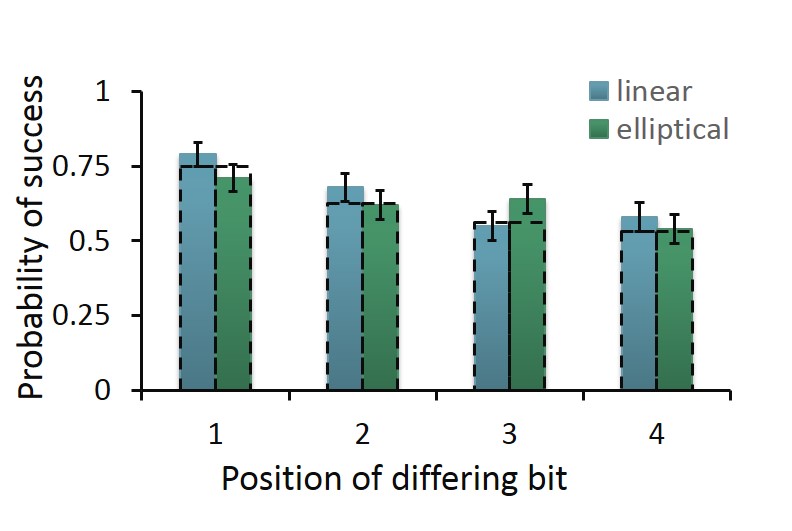}
\caption{Success probability for the Millionaires Problem using the linear (blue/first bar) and elliptical (green/second bar) scheme. We compare different four-bit numbers in binary representation to the number $0101$, where each compared number deviates from $0101$ in exactly one bit. The expected probability of success (shown in black dotted lines) depends on the position of the differing bit (1 corresponds to the most significant bit, 4 to the least significant bit). Error bars derive from binomial statistics. }
\label{Yao} 
\end{figure}

The first class we consider is a program built from a combination of $\mathcal{G}_2$ gates which are universal for classical computation. We use it to solve Yao's Millionaires Problem \cite{Yao82}, in which two people wish to compare their wealth without disclosing this value to the other party. To accomplish this goal, Alice encodes her wealth into the program. Bob's wealth will be his input (see \fig{millionaires} of the Supplementary Information). The program returns a single bit, indicating which number is larger. We ran the Millionaires Problem using both the linear and the elliptical schemes on several inputs. Alice encoded a four-bit number and Bob compared it to numbers that each differed in one bit from Alice's input. The detailed results are shown in \fig{Yao}. In good agreement with our theoretical expectations, it can be seen that the probability of success rises with the significance of the bit in which the two numbers differ (i.e. it is easier to discriminate two numbers that differ in the most significant bit than two that differ in the least significant bit). 

\begin{figure}[h]
\centering
\includegraphics[width=0.5\textwidth]{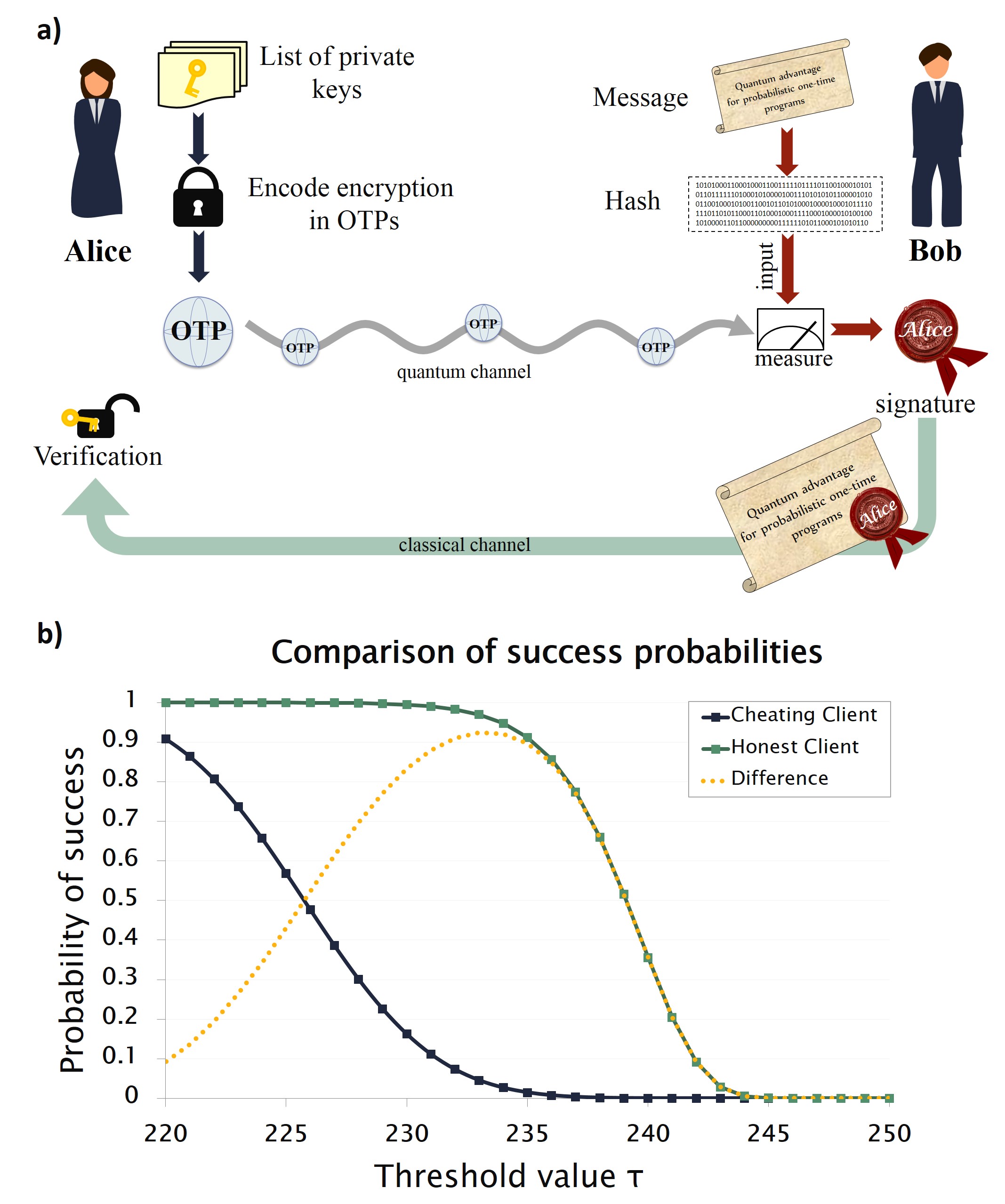}\caption{Private key  one-time signature scheme. a) Overview of the private key one-time  signature scheme. Alice encodes her signature program as noisy quantum OTPs, and sends them to Bob via a quantum channel. Bob measures the states according to the hash of a message he wishes to sign, with outputs corresponding to a signature. This message and signature pair may be verified by Alice at a later stage. b) Bob's success probabilities of signing a message or messages, in both the honest (one message) and the dishonest (two messages) case. The probability of the client successfully generating a valid signature is plotted against the threshold value for number of correct bits required to sign a message, given that the client is provided with 300 OTPs for each row of the table. The green line shows the probability of an honest client generating a single signature ($\geq \tau$ correct outputs for each row of the signature), as a function of the threshold number of correct bits required in each row. The blue line shows probability of a dishonest client generating two signatures, which differ only by one bit in the hash (the probability of a client following an honest strategy in all but one row). The difference between the probabilities of the prior two lines is indicated by the yellow line. Details are given in the Supplementary Information. }\label{DigitalSignature}
\end{figure}

The second kind of program we consider concerns the delegation of digital signatures or one time power of attorney. Here Alice can enable Bob to sign one, and only one, message of his choice with a signature derived from her private key. However, due to the probabilistic nature of the described OTPs, there is a non-negligible probability that OTPs will not output the correct signature. To compensate for this we may repeat the procedure and define some threshold number of signatures which is announced publicly to be an acceptable number required to verify a given message has been signed. Alice produces many distinct OTPs each using a different private key such that Bob has a high probability of forming the required number of signatures for a single message.

Standard signature schemes use a public key for verification \cite{Schneier1993}. However, due to technical reasons limiting our gate rate in experiment, we restrict  our demonstration to a symmetric digital signature scheme, wherein Alice's private key is used to verify a signature. Such a program may be utilised for a third party to spend an amount of money on someone else's behalf, so that they should pay anyone with a signed receit. An overview of this scheme is shown in Fig. 5a.  Bob computes a hash of the message he wishes to sign and uses this as the input to the OTPs (using a hash ensures that the input length does not depend on the length of the actual message signed). The output of the OTPs will then be the digital signature which Alice may verify. For each bit of this hash Alice provides 300 $\mathcal{G}_1$ gates, from which Bob produces a bit string dependent on the result of measuring according to that bit. Such a bit string may be compared by Alice to the ideal case where all gates have been implemented on the corresponding hash bit. We require that each bit string matches  such an ideal string in at least $\tau$ positions to produce a valid signature. The threshold $\tau$ is chosen as a function of the bit string length $T$ to maximise the difference between the probabilities of success of the honest and dishonest strategies, wherein a dishonest strategy Bob would attempt to sign two hashes differing only by a single bit. This is illustrated in \fig{DigitalSignature}b and in \fig{histogram} of the Supplementary Information. As $T$ is increased the probability of an honest Bob forming a sufficient fraction of correct bits ($\geqslant \tau/T$) in each bit string approaches 1, while that of a dishonest user who would try to form multiple signatures, approaches $0$. This demonstrates a clear example of a case where even probabilistic one-time programs enable new functionality that is inexecutable using classical technology.

\section*{Security analysis}

We will now discuss the security of our protocol and show a strict advantage over any possible classical strategy. We note that the security relies on several measures affecting  different steps of our protocol. Starting with the logical synthesis we see that when  gate-OTPs are combined into circuits, there is some freedom over how the gates are chosen. In our proof-of-principle demonstration of Yao's millionaires problem we limit the information accessible to Bob by randomly inserting  pairs of NOT gates into the circuit immediately after each gate-OTP with probability one-half. The first NOT gate is absorbed backwards into the gate-OTP, altering the encoded gate. The second NOT of the pair is propagated forward, through any present fan-out and XOR gates, and absorbed into the next layer of gate-OTPs, altering the function they encode. Such a procedure can always be applied to any circuit composed of gate-OTPs along with XOR, NOT and fan-out operations.

To analyse the effect of this randomisation procedure, we will assume it is applied after every gate-OTP. In such a case, the joint state of the quantum systems used to encode the gate-OTPs is maximally mixed, and hence independent of the encoded function. For those gate-OTPs which produce the output of the  program the second NOT gate cannot be absorbed into a subsequent gate-OTP. We will simply eliminate this second NOT gate, effectively applying a one-time pad to the program's output and creating the maximally mixed state from the perspective of the receiver. Such a scheme thus negates all losses in the system as the maximally mixed state does not allow a dishonest user to extract any information regarding the intended gate-OTP.  Since the output of the program can be revealed by decoding the one-time pad, the accessible information for the entire system can be no greater than the size of this encryption key, and hence can be no greater than the number of output bits for the program. This is in line with the requirement that a one-time program should reveal no more information than can be obtained from a single run of the program.

\begin{figure}[t]
\centering
\includegraphics[width=0.48\textwidth]{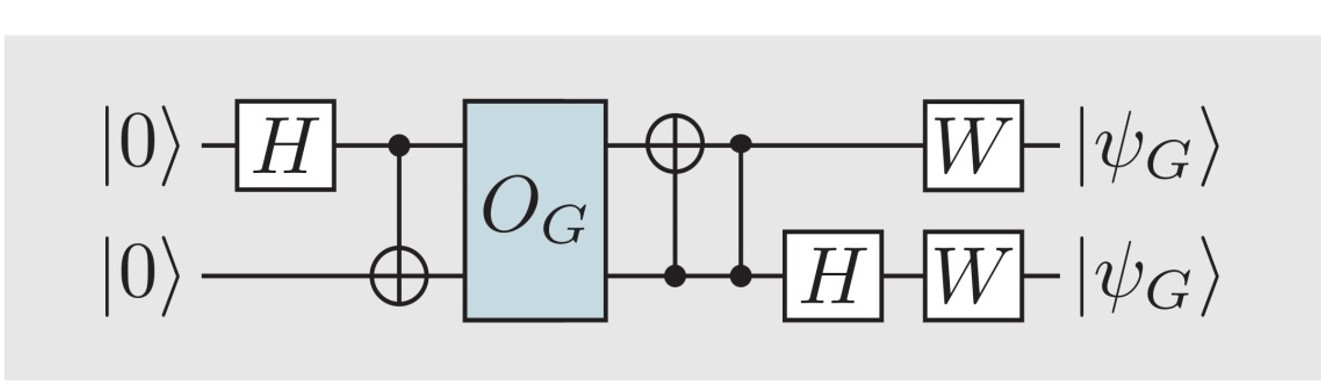}\caption{ Security of probabilistic one-time programs. The quantum circuit produces two copies of the state $\ket{\psi_G}$ for any $G\in \mathcal{G}_1$. Here $O_G$ implements a reversible version of the encoded gate $G$ mapping $\ket{x,y}$ to $\ket{x,y\oplus G(x)}$ and  $W=\left[c,s;s,-c\right]$, where $c=\cos(\pi/8)$ and $s=\sin(\pi/8)$.}
\label{Security}
\end{figure}

We now consider the security of the individual gate-OTPs corresponding to gates in $\mathcal{G}_1$. We show that strictly less can be learned from a single copy of them than from a single query to the encoded function (i.e. an ideal one-time implementation of that function). For all gates $G \in \mathcal{G}_1$, the corresponding state $\rho_G$ is pure, and so we will denote the state vector as $\ket{\psi_G}$. 
\fig{Security} shows how a single query of the encoded function can be used to produce two copies of this state.  
The fact that states encoding different programs are non-orthogonal, coupled with the no-cloning theorem \cite{Wooters1982, Dieks1982}, implies it is not possible to produce two copies of $\ket{\psi_G}$ from a single copy, and hence strictly less can be learned about $G$ from a single copy of $\ket{\psi_G}$ than from a single (coherent) query to the function it encodes.

We conclude our analysis by discussing the security of $\mathcal{G}_1$ and $\mathcal{G}_2$ gates. We show that the gate-OTPs we have explored here have strict advantages over any purely classical computational procedure. First, we choose an appropriate figure of merit for which to compare quantum and classical noisy OTPs. An ideal OTP would allow for one, and only one, evaluation of the encoded function, resulting in exactly one input-output pair. We will therefore choose the average probability of evaluating a specific input-output pair correctly, $P_1$, compared to the average probability of correctness when evaluating all input-output pairs $\tilde{P}_1$. In the classical case information can always be copied. Therefore, a classical procedure producing one input-output pair with some fixed probability of success can be repeated arbitrarily many times to produce a noisy version of the encoded gate. The probability of getting a specific input-output pair is equal to the average probability across all input-output pairs, thus $P_1^{C}=\tilde{P}^{C}_1$. However, for $\mathcal{G}_k$ OTPs this is not the case. If we fix the single line probability of success such that $P_1^{C}=P^{Q}_1$ we find that, for $\mathcal{G}_1$ gates $\tilde{P}_1^{Q}=0.75$ while $\tilde{P}_1^{C}\approx 0.8536$. Similarly, for $\mathcal{G}_2$ we find that $\tilde{P}_1^{Q}=0.625$ while $\tilde{P}_1^{C}=0.75$. This shows that our encoding gives an advantage over the best possible classical scheme for an equivalent $P_1$. Details of these calculations can be found in the Supplementary Information, where it is also shown that success probability can be boosted via error-correction while still maintaining an advantage. Furthermore, in the case of $\mathcal{G}_1$ gates, we may state that the probability of an adversary finding the parity of two lines, which gives an upper bound on the probability of guessing the complete truth table, is strictly lowers than in any possible classical encoding. This includes noisy implementations of oblivious transfer \cite{Rabin1981,Wullschleger2006} as our $\mathcal{G}_1$ gates are equivalent to noisy $\binom{1}{2}$-oblivious transfer. Remarkably, even though oblivious transfer with a vanishing error probability is known to be impossible \cite{Lo1997,Wullschleger2006}  with our digital signature scheme we were able to present an implementation whose overall success probability can approach $1$.

Aside from the inherent security of an ideal implementation of gate-OTPs, additional measures are necessary in the presence of communication over lossy channels. It is not in general advisable for Alice to simply resend qubits that are not received by Bob, since he can simply claim to have lost a photon to receive a new copy and hence gain additional information about the encoded gate-OTP. This may be prevented via a simple subroutine: for each gate several copies of each state are produced, but each with a randomly chosen additional one-time pad (i.e. a bit flip on the output of all possible inputs). These states are thus in the maximally mixed state as observed from the client and provide no information. Alice will reveal only the one-time pad for the state that Bob confirms to have received and that she wants him to use. Bob will then keep or flip his measurement result, according to Alice's one-time pad and proceed with the next gate following the same procedure. This procedure has been used in each of the demonstrated programs. 

\section*{Discussion}
Here we have shown the implementation of probabilistic one-time programs in theory and experiment. Our results demonstrate that quantum physics allows for better security trade-offs for certain secure computing tasks than are possible in the classical world, even when perfect security cannot be achieved. This is achieved without assumptions on computational hardness, noisy storage or difficulty of entanglement.   Using readily available technology we find  our results are in excellent agreement with the theoretical predictions. Future advances in technology that would allow for non-separable measurements on the client's side could be used to further improve our implementation.     
We believe the presented work strongly hints at a rich area of quantum protocols to enhance the security of classical computation, even before large-scale quantum computers can be realised.

\putbib[OTP_bib]
\end{bibunit}

\section*{Acknowledgements} We thank I. Alonso Calafell, M. Tillmann and J. Zeuner for helping with the electronics and L. Rozema, A. Sharma, T. Str{\"o}mberg and T. Withnell for discussions. M.-C.R. acknowledges support from the the uni:docs fellowship program of the University of Vienna. T.B.B. and P.W. acknowledge support from CAPES through the Science Without Borders program (grant PDSE 99999.005394/2014-07). J.F.F. and J.A.K. acknowledge support from the Singapore National Research Foundation (NRF-NRFF2013-01). J.F.F. and P.W. acknowledge support from the Erwin Schr\"{o}dinger Institut at the University of Vienna. J.F.F. acknowledge support from United States Air Force Office of Scientific Research (FA2386-15-1-4082) and P.W. acknowledges support from the Austrian Research Promotion Agency (FFG) through the QuantERA ERA-NET Cofund project HiPhoP; and the Austrian Science Fund (FWF) through START (Y585-N20) and the doctoral program CoQuS (No.W1210); and the United States Air Force Office of Scientific Research (FA9550-16-1-0004) and (FA2386-17-1-4011); and the Research Platform TURIS at the University of Vienna. The authors are named on a patent application relating to this method of implementing probabilistic one-time programs (application numbers EP16162886 and PCT/EP2017/057538).

\newpage

\clearpage

\section*{Supplementary Information}

\renewcommand\thefigure{\Alph{figure}} 
\setcounter{figure}{0}
\begin{bibunit}

\subsection*{Advantage over classical encodings for $\mathcal{G}_1$ gates}

We will show that the quantum implementation of $\mathcal{G}_1$ gates can hide more information than a classical scheme about the result of multiple lines of a truth table. This will be done by obtaining an inequality which must be satisfied by all classical schemes but is violated by the quantum scheme described in the main text.

We consider a situation where Bob is interested in the parity of some subset of lines of the truth table, which gives us a bound on the probability of identifying these lines in the subset exactly. Using equation 1 of the main text, he considers two states that are formed by summing over all states with equal parity over the subset of lines he is interested in. For subsets consisting of more than one line, these two states are equal and thus impossible to distinguish. In other words, while a single line of the truth table can be found with probability larger than $\frac{1}{2}$, the parity of two or more lines is completely hidden. For comparison, a classical scheme that encodes $\mathcal{G}_1$ gates with single-line error probability higher than $3/4$ must give correct results about the parity of two lines with probability higher than $1/2$. In particular, if the single-line error probability is $\frac{1}{2}+\frac{1}{2\sqrt{2}}$ (the same as achieved with the quantum states in equation 1 of the main text), the classical scheme must allow the parity of two lines to be correctly identified with probability at least $1/\sqrt{2}$, which is greater than $1/2$.

In order to improve the probability of getting the correct output from a particular gate, the programmer may send multiple copies, $c$, of the state corresponding to this gate. The client is expected to make some (possibly non-local) multi-qubit measurement to evaluate the line of the truth table corresponding to their input. In this case the statement that the parity of multiple lines of the truth table of the encoded gates are perfectly hidden is no longer valid. Sending multiple copies of the state in equation 1 of the main text creates a trade-off situation between precision and security, where precision is quantified by the success probability that an honest client can achieve when evaluating a single line, and security is quantified by the amount of information that can be found about multiple lines of the truth table simultaneously. A complete lack of security occurs when the client can perfectly identify which one of the gates is represented by the quantum state he possesses.

We will now make a comparison between a quantum scheme and what could be achieved by a classical scheme. Any classical scheme encoding a gate can be repeatedly rerun to generate a noisy truth table for the encoded gate. We will see that these noisy truth tables must satisfy an inequality that bounds the maximum level of security that can be achieved for a given level of precision. This inequality is violated in the quantum case, allowing us to achieve more security for the same level of precision than any classical scheme.

\subsubsection*{Analytical results - Classical scheme for $\mathcal{G}_1$ gates}

Without loss of generality, we consider a classical model in which a programmer introduces some errors in the gate truth table. These are errors purposefully introduced at compile time. There is no point in introducing random errors at run time, since the client can evaluate the truth table multiple times and find the most common value with high probability. If there is some anti-correlation in the presence of errors in different lines, then the probability of getting a second line correct is decreased when conditioned on getting the first line correct.

To obtain this anti-correlation we consider that the programmer introduces $h$ errors in the truth table (with $0\leq h \leq 2$) with probability $E_h$ (so $E_0 + E_1 + E_2 = 1$). If one error is introduced it can affect either line with equal probability. Thus for an honest client interested in a single line of the truth table, the average probability that the obtained result is correct is
\begin{align}
F^{\text{C}}_{1} = E_0 + \frac{1}{2} E_1
\end{align}
Meanwhile, for a dishonest client interested in the parity of both lines, the probability that the obtained result is correct is
\begin{align}
F^{\text{C}}_{2} = E_0 + E_2
\end{align}

We can invert these equations to find $E_h$ in terms of $F^{\text{\text{C}}}_{1}$ and $F^{\text{\text{C}}}_{2}$. This tells the programmer what is the probability distribution in the number of errors that need to be introduced in order to produce an encoding that is characterized by given values of the probability of decoding a single line and of decoding the parity of both lines. This leads to the result
\begin{subequations}
\begin{align}
E_0 &= F^{\text{C}}_{1} + \frac{1}{2} F^{\text{C}}_{2}  - \frac{1}{2} \\
E_1 &= 1 - F^{\text{C}}_{2} \\
E_2 &= \frac{1}{2} - F^{\text{C}}_{1} + \frac{1}{2} F^{\text{C}}_{2} 
\end{align}
\end{subequations}
Each of these terms must be non-negative, which is only possible if
\begin{align}
F^{\text{C}}_{2} \geq \left| 2F^{\text{C}}_{1} - 1 \right|
\label{eq:classical_inequality}
\end{align}

This means that an attempt at a classical noisy gate which outputs correct results with probability $F^{\text{C}}_{1}$ also allows one to probe the parity of both lines of its truth table with probability greater than $\left| 2F^{\text{C}}_{1} - 1 \right|$. However, we will see that a quantum implementation of the noisy OTPs violate this inequality, showing that it hides more information about other lines of the truth table than is possible classically.

\subsubsection*{Analytical results - Quantum scheme for $\mathcal{G}_1$ gates}

The figure of merit that we consider for security in this section is $F_h$, the probability of success in calculating the parity of a subset of the lines of the truth table, as a function of the size $h$ of this subset (with $1 \leq h \leq 2$ in the case of $\mathcal{G}_1$ gates, whose truth table has only two lines). The outcome of the parity determination is binary, so we can use known results on quantum state discrimination of two quantum states. Specifically, the optimal probability of distinguishing them is uniquely determined by the 1-norm of half of their difference. For an honest client who is interested in only the first line of the truth table, the probability of success is related to the 1-norm of the operator
\begin{align}
\hat{A}_{1}  =\frac{1}{4}\rho_{00}^{\otimes c}+\frac{1}{4}\rho_{01}^{\otimes c}-\frac{1}{4}\rho_{10}^{\otimes c}-\frac{1}{4}\rho_{11}^{\otimes c}
\end{align}
while for a dishonest client who is interested in obtaining the parity of both lines, the probability of success is related to the 1-norm of the operator
\begin{align}
\hat{A}_{2} =\frac{1}{4}\rho_{00}^{\otimes c}-\frac{1}{4}\rho_{01}^{\otimes c}-\frac{1}{4}\rho_{10}^{\otimes c}+\frac{1}{4}\rho_{11}^{\otimes c}
\end{align}

Gates with one bit of input may be encoded as pure states, so providing multiple copies of them does not increase the dimensionality of the effective Hilbert space, which is spanned by at most four linearly independent vectors. For simplicity, we consider the case where the number of copies is odd and obtain the following results
\begin{align}
F^{\text{Q}}_{1} & \equiv\frac{1}{2}+\frac{1}{2}\left\Vert \hat{A}_{1}\right\Vert _{1}=\frac{1}{2}+\frac{1}{2}\sqrt{1-\frac{1}{2^{c}}} \label{eq:optimaldistinguishing}\\
F^{\text{Q}}_{2} & \equiv\frac{1}{2}+\frac{1}{2}\left\Vert \hat{A}_{2}\right\Vert _{1}=\frac{1}{2}+\frac{1}{2}\sqrt{1-\frac{2}{2^{c}}}
\label{eq:optimaldistinguishingparity}
\end{align}
where the superscript $\text{Q}$ refer to a quantum implementation.
These values do not satisfy the inequality in Equation \ref{eq:classical_inequality}. This means that, if we compare a classical scheme which offers the same level of precision for an honest client (i.e., $F^{\text{C}}_1 = F^{\text{Q}}_1$), the probability of success for a dishonest client is higher in the classical case. Similarly, if we restrict the two protocols to the same level of security as quantified by the probability of finding the parity of both lines, then the quantum OTP can offer better performance for honest clients than any classical scheme.

\subsection*{Advantage over classical encodings for other gates}

 We demonstrate the advantage of the quantum one-time programs with multiple bits of input over possible classical schemes. We assume that the gate-OTP is \textit{a priori} equally likely to encode any of the possible gates in $\mathcal{G}_k$. Although we will focus on $\mathcal{G}_2$ gates, parts of this discussion can be generalized to $\mathcal{G}_k$ gates with $k>1$.

We consider the probability distribution for (potentially correlated) Bernoulli random variables $X_i$, which are equal to 1 if and only if a query to the a noisy classical truth table encoding gate $G$ for input $i$ returns $G\left(i\right)$. 
All probability distributions over truth tables can be described in this way, and so it can be used to obtain a bound on the trade-offs inherent in any classical scheme. The sender does not know in advance which lines of the truth table the client might be interested, thus his/her interest is in minimizing the worst-case probability of correctly obtaining the output for multiple lines across all sets of lines. To do that, every line is treated in an equivalent way, and so all the elements on the diagonal of the covariance matrix of the Bernoulli variables $X_i$ will be equal, as will be all off-diagonal elements. The covariance matrix thus has the form $u\mathbb{I} - v \mathbb{M}$, where $\mathbb{I}$ is the identity matrix and $\mathbb{M}$ is the matrix with all entries equal to one. In order to obtain a fixed probability $P_1$ of correctness for a single query to a line of the truth table, it must be the case that $u-v = P_1 - P_1^2$.  Furthermore, since the minimum eigenvalue of such a matrix is $u - 2^k v$ and covariance matrices are positive semi-definite, it must be the case that $u - v \geq (2^k-1)v$. With these arguments, it's possible to bound the probability of obtaining the correct values for two lines (indexed by $x$ and $y$, with $x\neq y$) of such a truth table,
\begin{align}
	\tilde{P}_2 &= E(X_x X_y) \nonumber \\
	&= E((X_x-P_1)(X_y-P_1)) + P_1^ 2\nonumber\\
	&= -v + p^2\nonumber\\
	&\geq -\frac{(u-v)}{2^k-1} + P_1^2\nonumber\\
	&=  \frac{P_1^2-P_1}{2^k-1} + P_1^2\nonumber \\
	&= \frac{2^k P_1^2 -P_1 }{2^k-1} 
\end{align}
As the probability of evaluating a single line of a $\mathcal{G}_2$ OTP is $P_1 = 0.75$, a noisy classical truth table with the same success probability gives $\tilde{P}_1 = 0.75$ across all lines. The probability of correctly decoding a pair of lines is at least $\tilde{P}_2 = 0.5$, independently of the chosen pair of lines.

We may now compare this to the average probability of finding the output values of encoded gates for pairs of inputs. In the quantum case, if the client is interested in a particular line of the truth table, it's possible to implement a quantum measurement strategy that is specifically tuned to increase the probability of getting this value correctly. However, this degrades the available information about the other lines. Thus, in a marked difference to the classical scheme, the probability $P_1$ of finding the correct value of a particular line is different (and higher) than the average probability $\tilde{P}_1$ of getting a line correctly when the client is trying to identify the whole truth table.
The same argument is valid when the client is interested in a given pair of lines, as compared to the average probability of correctly identifying pairs of lines when trying to identify the whole truth table.

Using the quantum encoding without error correction, a client interested in a given line of the truth table of a $\mathcal{G}_2$ gate can correctly identify it with probability equal to $0.75$. On the other hand, when the client tries to identify the whole truth table, the average success probability is only $0.625$. Looking at pairs of lines, making a specific measurement can allow the client to obtain a probability of success equal to $0.5$, but the average over all lines in a measurement of all lines is only $0.375$.

\subsection*{Optimal measurements}

We turn our attention to the measurement strategy that a dishonest client could follow if he is interested in identifying all lines of the truth table of an encoded gate $G$. This problem is cast as a quantum state discrimination of one state among $2^k$ alternatives, and the figure of merit is the probability of making a correct guess about the entirety of the truth table. We will consider the \textit{pretty good measurement} strategy introduced by Hausladen and Wootters \cite{hausladen1994} and another strategy introduced by Je{\u z}ek, Reh{\'a}{\u c}ek and Fiur{\'a}{\u s}ek \cite{jevzek2002}. Because of some properties of the way that the gates are encoded in quantum states, these two strategies are equal and also optimal, because they obey established criterias \cite{Bae15}.

Considering an ensemble of states $ \lbrace \rho_i \rbrace$ with a priori distributions $\lbrace q_i\rbrace$, there are several good measurement strategies for distinguishing them. One candidate strategy is the \textit{pretty good measurement}, or $\mathcal{PGM}$ \cite{hausladen1994}, for which the POVM correspoding to an output $x$ (where $x$ is a $2^k$-bit string representing the truth table) is
\begin{align}
M_x^{\mathcal{PGM}} = \left( \sum_{s=1} q_s \rho_s \right)^{-1/2^+} q_x \rho_x \left( \sum_{l=s} q_s \rho_s \right)^{-1/2^+}
\end{align}
where the operation $A^{-1/2^+}$ is defined as
\begin{align}
A^{-1/2^+} = \sum_{j : a_j >0} a_j ^{-1/2} \ket{a_j}\bra{a_j}
\end{align}
 with $a_j$ and $\ket{a_j}$ being the eigenvalues and eigenvectors of $A$. In a similar manner, slightly more complex sets of measurement operators may be formed that are called the Je{\u z}ek-Reh{\'a}{\u c}ek-Fiur{\'a}{\u s}ek iterative measurement operators \cite{jevzek2002}. These are defined recursively, where each iteration is indexed by $w$. 
\begin{align}
M_x^{\mathcal{JRF},w} &= \left( \sum_{s=1} q_s^2 \rho_s M_s^{(w-1)} \rho_s  \right)^{-1/2^+} q_x^2 \rho_x M_x^{(w-1)}  \rho_x \nonumber \\ &\times  \left( \sum_{s=1} q_s^2 \rho_s M_s^{(w-1)} \rho_s \right)^{-1/2^+}
\end{align}
where in the first iteration $M_x^{\mathcal{JRF},0}= \mathbb{I} / 2^k$.

Due to the form of our states, each satisfy $\rho_x^2 = \xi \rho_x ,~ \forall x$, where $\xi$ is a proportionality constant independent of $x$. Coupled with an assumption that the states are \textit{a priori} equiprobable (so that $q_x = \zeta ,~ \forall x$, where $\zeta$ is a constant independent of $x$), the $\mathcal{PGM}$ operators  are equal to every iteration of the $\mathcal{JRF}$ operators.

It is known that a POVM strategy is optimal if it satisfies the following two conditions \cite{Bae15},
\begin{align}
M_x \left( q_x \rho_x - q_y \rho_y \right)  M_y &= 0 ~,  \forall x,y
\\
\sum_{x=0}^{2^k-1} q_x \rho_x M_x - q_y \rho_y  &\succeq 0 ~,  \forall y
\end{align}
In a numerical study, we have verified that the $\mathcal{PGM}$ strategy is optimal when $\rho_x$ represents three or less copies of the gate-encoding states.

\subsubsection*{Optimality of the measurement for single copies}

We now give an analytical proof that the $\mathcal{PGM}$ strategy is optimal when a single copy of the quantum states is sent. First we note that, in the case of a single copy, $ \sum_{s} q_s \rho_s = \mathbb{I}$, assuming as before that the states are \textit{a priori} equiprobable (when the number of copies is larger than one, then $\sum_{s} p_s \rho_s \neq \mathbb{I}$ even in the equiprobable case). Thus, the measurement operators $M_x$ are proportional to the density matrices $\rho_x$.

We will now obtain a bound on the value of $\text{tr}\left(\rho_x Mxi\right)$ using H{\"o}lders inequality,
\begin{align}
\Vert fg \Vert _1 \leq \Vert f \Vert_p \Vert g \Vert_q
\end{align}
which is valid when $\frac{1}{p}+\frac{1}{q} = 1$. Assuming that each state is equally probable and the normalization condition $\sum_x M_x = \mathbb{I}$, this implies that $\text{tr}\left(M_x\right) =D/2^k ~, \forall x$. We also note that $M_x \succeq 0$ as required for POVMs. Using the values $p=1$, $q = \infty$, $f=M_x$ and $g=\rho_x$ in H{\"o}lders inequality, we find that
\begin{align}
\left| \text{tr}( M_x \rho _x) \right| &=  \left\Vert  M_ix\rho_x \right\Vert _1 \nonumber \\
&\leq  \left\Vert  M_x \right\Vert_1 \left\Vert \rho _x \right\Vert_\infty  \nonumber \\
&=  \frac{D}{2^k} \frac{2}{D}\nonumber \\
&= 2^{1-k}
\end{align}
We now use the POVM given by the $\mathcal{PGM}$ (or $\mathcal{JRF}$) operators and show that this saturates the above inequality.
\begin{align}
\text{tr}(M_x \rho_x ) &= \sum_j u_j r_j \nonumber \\
&= \sum_{j : u_j, r_j \neq 0} u_j \frac{2}{D} \nonumber \\
&= \text{tr}(M_x)\frac{2}{D} \nonumber \\
&= \frac{D}{2^k} \frac{2}{D} \nonumber \\ 
&= 2^{1-k}
\end{align}
where $u_j$ and $r_j$ are the eigenvalues of $M_x$ and $\rho_x$ respectively and we have used the fact that, because $M_x \propto \rho_x $, they both have degenerate eigenvalues in the same eigenbasis. As the $\mathcal{PGM}$/$\mathcal{JRF}$ measurement achieves the exact upper bound on $\vert \text{tr}(\rho _x M_x) \vert$, they must be optimal. 
 
\subsubsection*{Applying the optimal measurements}

We now look at situation where the client uses these operators to try to learn what state was sent. We may quantify the number of lines of the truth table the client can on average obtain correctly. We define $E_h$ as the probability that \textit{exactly} $h$ lines are incorrect; in other words, $h$ is the Hamming distance between the encoded gate and the result of a measurement that tries to identify the gate.  We consider the average taken over all gates in $\mathcal{G}_k$, and hence over all $\rho_x$,
\begin{align}
E_{h}=\sum_x q_x \sum_{s : \mathcal{H}(s,x) =h} \text{tr}\left(\rho_{x} M_{s} \right)
\end{align}
where $\mathcal{H}\left(s,x\right) $ is the Hamming distance between the truth tables represented by $s$ and $x$. From this average number of errors, we can then consider the average probability of correctly identifying a subset of $L$ lines, $\tilde{P}_L$, which is given by
\begin{subequations}
\begin{align}
\tilde{P}_1 &=E_0 + \frac{{{3}\choose{1}}}{{{4}\choose{1}}}E_1 + \frac{{{2}\choose{1}}}{{{4}\choose{1}}}E_2 + \frac{{{1}\choose{1}}}{{{4}\choose{1}}}E_3 \nonumber \\ &= E_0 + \frac{3}{4}E_1 + \frac{1}{2}E_2 + \frac{1}{4}E_3  \\
\tilde{P}_2 &=E_0 + \frac{{{3}\choose{2}}}{{{4}\choose{2}}}E_1 + \frac{{{2}\choose{2}}}{{{4}\choose{2}}}E_2 = E_0 + \frac{1}{2}E_1 + \frac{1}{6}E_2 \\
\tilde{P}_3 &=E_0 + \frac{{{3}\choose{3}}}{{{4}\choose{3}}}E_1 = E_0 + \frac{1}{4}E_1 \\
\tilde{P}_4 &= E_0
\end{align}
\end{subequations}
In \fig{p1vsp1} of the Supplementary Information, $\tilde{P}_1$ in the quantum case is plotted against $\tilde{P}_1$ in the classical case, which is simply the probability that a single line is correct, for $\mathcal{G}_2$ gates. This shows a clear quantum advantage for noisy one time programs.

\subsection*{Description of the Private Key Signature scheme}

This scheme allows Alice to delegate to Bob the power of digitally signing a message of his choice once and only once. To realize this, Alice's digital signature will be formed by the output of one-time programs. These OTPs take Bob's message as an input and output a valid signature. To allow the signing algorithm to work on a fixed-size input Bob creates a hash of his message using SHA3-224 protocol (there is no particular theoretical reliance on this or any particular hash, but we chose to use SHA3-224 in our demonstration).  The signature is verified by Alice, the programmer, by comparing the generated signature against the ideal one that would be produced in the case of perfect OTPs. For each bit of the hash the client is provided with $T$ OTPs, each of which is chosen uniformly at random from the set of $\mathcal{G}_1$ OTPs (in principle we could use $\mathcal{G}_k$ gates, but we chose to use $k=1$ in our demonstration). The client makes measurements on these states according to the corresponding bit of his hash, producing an array where each row corresponds to the output bits for a single hash bit. The signature is deemed to pass if each row is correct in at least $\tau$ places, wherein the threshold $\tau$ is a integer predetermined by the programmer. We will show now how the scheme displays a clear example of a situation where even probabilistic OTPs may be used to implement a program which works with a high probability of success. 

We compare the probability of success of 	passing the verification step for an honest client signing one message to the probability of passing the verification step twice for a dishonest client signing two messages which hash to different values. We will consider the cases where the hashes differ by only one bit. This is a worst case scenario in which an adversary has the maximum probability of cheating successfully. The threshold value $\tau$ is chosen to maximise the difference between the success probabilities for an honest and a dishonest client in such a case.
\begin{center}
\textit{Probability that a dishonest client can pass the verification step for a single bit of the hash:}
\end{center}
The two signatures taken together constitute a string of lenght $2T$. Each signature needs to be correct in at least $\tau$ places to pass the verification stage and thus a necessary (but not sufficient) condition for the combined string to pass is that it matches the concatenation of the two ideal signatures in $2\tau$ places.
We place an upper bound on the probability of this happening by using a similar method to that used by Vazirani \cite{Umesh}. The two ideal signatures are encoded in $T$ qubits as is the case when we are sending $T$ $\mathcal{G}_1$ OTPs. It's considered that each of the $2T$-bit strings corresponding to possible signatures is mapped to a pure state $\ket{\phi_x}$, while a measurement that would output a $2T$-bit string $y$ is associated with a projector $P_y$. This can be done without loss of generality since the measurement projectors can be defined in a larger Hilbert space than the received OTP state, since $\ket{\phi_x}$ may contain an arbitrary number of additional ancilla qubits. The probability that at most $h$ mistakes are made in such a decoding protocol is given by
\begin{align}
\mathcal{P} &\equiv \text{Prob}\left( H(x,y) \leq h \right) \nonumber \\ &= \frac{1}{2^{2T}} \sum_{x,y : H(x,y) \leq h} \text{tr}\left(P_y \left| \phi_x \right\rangle\!\left\langle \phi_x \right| \right)
\label{eq:probability_Hamming}
\end{align}
where $H(x,y)$ is the Hamming distance between the strings $x$ and $y$.

At this moment it is helpful to analyse some properties of the specific ways in which the $\ket{\phi_x}$ states are defined. The $2T$ bits of $x$ are split in pairs corresponding to the $i$-th bit of each signature, and each pair is encoded in a qubit using the model for $\mathcal{G}_1$ gate-OTPs. Thus, all $\ket{\phi_x}$ states can be written as
\begin{align}
\ket{\phi_y} &= \ket{\phi_{y_1}} \otimes \ket{\phi_{y_2}} \otimes \cdots \otimes \ket{\phi_{y_T}} \otimes \ket{\mathcal{A}} 
\end{align} 
where $y_k \in \{00,01,10,11\}$ and $\ket{\mathcal{A}}$ represents the state of an arbitrary-dimensional ancilla, which does not depend on $y$.
Two states $\ket{\phi_x}$ and $\ket{\phi_y}$ are orthogonal if there is at least one pair of bits (which are encoded in the same qubit) which differ between $x$ and $y$ in both bits. This suggests a way to find an orthonormal basis for this space, by starting with any $\ket{\phi_y}$ and obtain other states by negating pairs of bits from $y$. Since there are $T$ pairs to negate and all states obtained this way are orthogonal to each other, they form a orthonormal basis with $2^T$ elements. Given that the space spanned by possible OTP states is of dimension $2^T$, and that every state can be written as a linear combination of some others, this basis must span the space generated by all $\ket{\phi_y}$ states. We call this the $y$-basis.

Using these properties, we argue that that the operator $\sum_{x : H(x,y) \leq h} \left| \phi_x \right\rangle\!\left\langle \phi_x \right|$ is diagonal in the $y$-basis just defined, and that $\ket{\phi_y}$ is the eigenvector corresponding to its largest eigenvalue. To see this, we need to consider what the strings $x$ appear in the sum. Specifically, for each string $x$ where the first bit of a given pair does not match the corresponding bit in $y$ (but the second bit of that pair does match), there is also another string where the first bit matches but the second bit does not match. These strings are always both included or both excluded, because the Hamming distance between each of them and $y$ is the same. The mixture associated with these two states is diagonal in the $y$ basis, even though none of them are individually.

With the eigenvectors already found, the task is to find eigenvalues. For an eigenvector $\ket{\phi_z}$, the eigenvalue depends on how many strings $x$ that are not orthogonal to $z$ are included in the summation. Because the summation over $x$ is centered around $y$ (in the sense of the Hamming distance), the eigenvector $\ket{\phi_y}$ has the highest number of strings $x$ appearing in the sum. This leads to this eigenvalue being the highest one. By a counting argument, it's possible to arrive at its specific value.  Strings $x$ that appear in the sum are at Hamming distance at most $h$ from $y$, but if both bits of a given pair are different in $x$ and $y$ then the state corresponding to this string does not contribute. 
If a pair is equal in $x$ and $y$, then the contribution to $\left\langle \phi_y \mid \phi_x\right\rangle\left\langle \phi_x \mid \phi_y\right\rangle$ corresponding to that qubit is $1$. If a pair has $x$ and $y$ differing in one bit, the contribution to $\left\langle \phi_y \mid \phi_x\right\rangle\left\langle \phi_x \mid \phi_y\right\rangle$ is $1/2$, but because there are two of those states, their sum also constributes $1$. Thus, when a Hamming distance of $w$ between $x$ and $y$ is considered, we must consider only terms where there is either zero or one differences per pair, with each configuration contributing 1. The eigenvalue corresponding to $\ket{\phi_y}$ is then
\begin{align} 
\lambda = \sum_{w=0}^h \left( \begin{array}{c} T \\ w \end{array} \right)
\end{align}
This was explicitly checked for small values of $T$ by numerical diagonalization.

We can now find an upper bound to the probability $\text{Prob}\left( H(x,y) \leq h \right)$ that a dishonest client can make at most $h$ mistakes in the determination of the $2T$-bit string corresponding to the ideal signatures for two distinct messages. Continuing from Equation \ref{eq:probability_Hamming}, we have that
\begin{align}
\mathcal{P}
&= \frac{1}{2^{2T}} \sum_y \text{tr}\left(P_y \sum_{x:H(x,y)<=h} \left| \phi_x \right\rangle\!\left\langle \phi_x \right|\right)
\nonumber \\ 
&\leq \frac{1}{2^{2T}} \sum_y \text{tr}\left(P_y Q\right) \sum_{w=0}^h \left( \begin{array}{c} T \\ w \end{array} \right)
\end{align}
where $Q$ is a projector to the codespace spanned by the codewords $\ket{\phi_x}$, which has dimension $2^T$. Then,
\begin{align}
\mathcal{P}
&\leq \frac{1}{2^{2T}} \text{tr}\left( \left(\sum_y  P_y\right) Q\right) \sum_{w=0}^h \left( \begin{array}{c} T \\ w \end{array} \right)
\nonumber \\ &=
 \frac{1}{2^{2T}} \text{tr}\left(Q\right) \sum_{w=0}^h \left( \begin{array}{c} T \\ w \end{array} \right)
\nonumber \\ &=
 \frac{1}{2^{T}} \sum_{w=0}^h \left( \begin{array}{c} T \\ w \end{array} \right)
\end{align}

If $h/T < (1/2) - \epsilon$, for any positive constant $\epsilon$, 
the probability of obtaining an output string within Hamming distance $h$ of the ideal signature string is exponentially small in $T$. Returning to the definition of $h$ as $2T-2\tau$, we see that the exponential suppression happens when the ratio $\tau/T$ is fixed as any constant greater than $3/4$. 
We now have an upper bound for the probability of success of a dishonest client passing the verification step for a single bit of the hash for two different inputs. As we assume a worst case scenario, where the hashes differ in only a single bit the client can follow the honest scenario for all other bits of his hash. Therefore, the overall probability of a dishonest client to sign two such messages is simply given by the product of the individual success probabilities per bit.

It becomes increasingly unlikely that the client is able to sign two messages if the required threshold for signing one message is set as a constant fraction $\alpha>3/4$ of $T$. If the threshold $\tau$ is set at lower than $\left(\frac{1}{2}+\frac{1}{2\sqrt{2}}\right)T \approx \left(0.85\right)\cdot T$, the honest client is able to sign a single message with probability that approaches 1 as $T$ is increased.

In conclusion, when the threshold $\tau$ is chosen to lie between $(0.75)\cdot T$ and $(0.85)\cdot T$, a client can sign one message with high probability but can sign two messages with low probability. In the limit of high $T$, these probabilities tend to 1 and 0, respectively. For practical reasons, as a trade-off between security and speed, we chose the values $T=300$ and $\tau=234$, which results in a client being able to sign one message with probability $97\%$, but with a smaller than $4\%$ probability of signing two messages which hashes to strings differing in only one bit. This is an upper bound to the cases where the hashes are different in more than one bit. Another interesting feature of the protocol is that it does not require a perfect implementation of the quantum states. Noise can be tolerated as long as the probability of obtaining a correct outcome for a single line of the $\mathcal{G}_1$ OTP is higher than $75\%$, provided that $\tau$ is chosen accordingly and $T$ is high enough such that the client can sign one message with reasonably high probability.

\section*{Supplementary Figures}

\begin{figure}[H]
\centering
\includegraphics[width=0.5\textwidth]{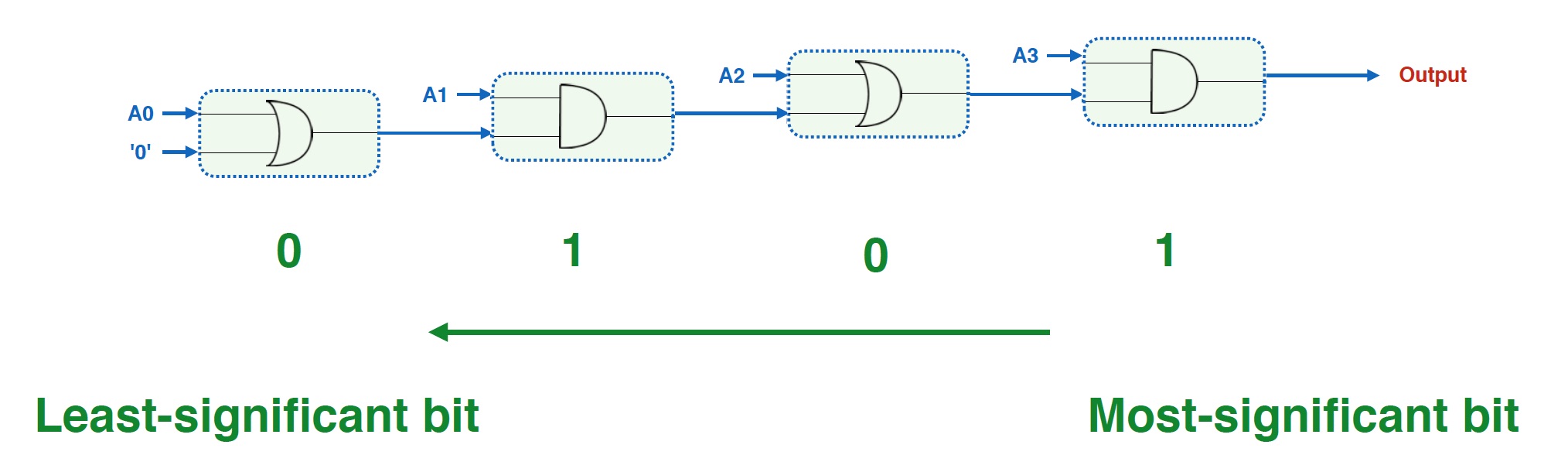}
\caption{Circuit diagram for the implemented solution to the Millionaires Problem. Alice encodes her input in binary representation by sending an OR gate for each bit with the value $0$ and an AND gate for each bit with the value $1$. } \label{millionaires} 
\end{figure}

\begin{figure}[H]
\centering
\includegraphics[width=0.5\textwidth]{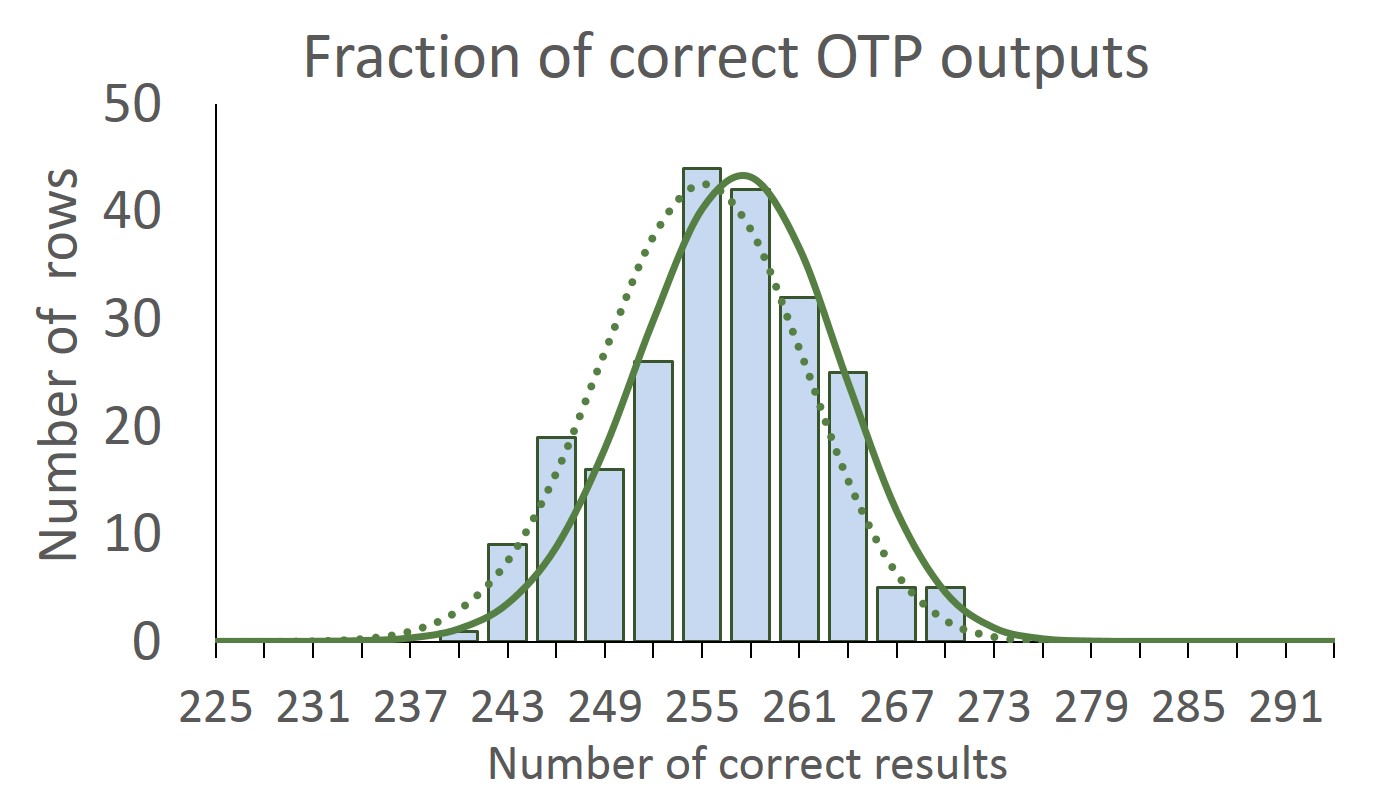}
\caption{The histogram presents the result of the experimental implementation of a delegated digital signature scheme, showing the distribution of number of correct output bits in each row compared to the theoretically expected values when a row length of 300 was used. In the experimental implementation the threshold $\tau = 234$ was chosen as this value maximises the difference between the success probabilities of the honest and dishonest client. It can be seen that this threshold was passed for every row. Considering the expected deviation due to the probabilistic nature of the scheme and experimental imperfections, the results are in good agreement with the theoretical predictions.
Solid green line: the binomial distribution wherein each $\mathcal{G}_1$ OTP has the theoretically expected success probability for a perfect implementation of probability of correctly giving the correct output.
Dotted green line: the binomial distribution based around the average probability of success that could be realised in the experiment (this being slightly reduced compared to the theoretical prediction due to experimental imperfections). Histogram bars are of width 3, with values taken from a single evaluation of a signature.}
\label{histogram} 
\end{figure}

\begin{figure}[H]
\centering
\includegraphics[width=0.5\textwidth]{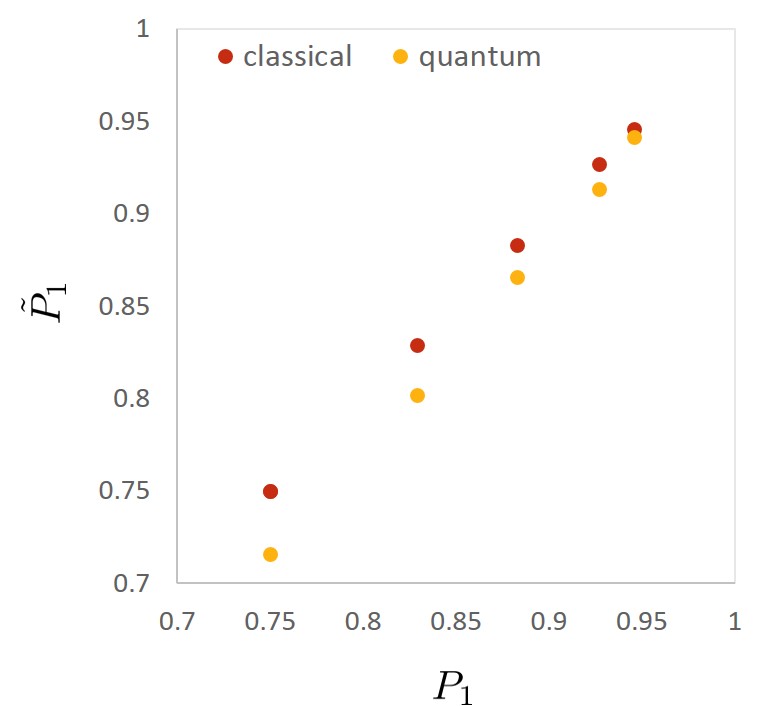}
\caption{The average probability of success when trying to find all lines of a truth table in the quantum case $\tilde{P}_1$ is plotted against the probability of finding exactly one line, $P_1$, and compared to classical case. The different probabilities of success for a single line could in the quantum case be achieved by using error correction as described in the Supplementary Information. This shows the probability that a single line is correct, for $\mathcal{G}_2$ gates and demonstrates a clear quantum advantage for the noisy one time programs.} \label{p1vsp1} 
\end{figure}

\clearpage

\section*{Supplementary Tables}

\begin{table}[h]
\resizebox{0.48\textwidth}{!}{
\setlength{\tabcolsep}{1pt}
\captionsetup[subfloat]{labelformat=empty}
\subfloat[]{
\begin{tabular}{@{}c@{}@{}c@{}}
\hline 
Gate & Encoding states \\
\hline 
0000 & $\begin{cases}
\left|\Psi_{0}\right\rangle \otimes\left|\Psi_{0}\right\rangle \otimes\left|\Psi_{0}\right\rangle \\
\left|\Psi_{0}\right\rangle \otimes\left|\Psi_{1}\right\rangle \otimes\left|\Psi_{Id}\right\rangle \\
\left|\Psi_{1}\right\rangle \otimes\left|\Psi_{0}\right\rangle \otimes\left|\Psi_{not}\right\rangle \\
\left|\Psi_{1}\right\rangle \otimes\left|\Psi_{1}\right\rangle \otimes\left|\Psi_{1}\right\rangle 
\end{cases}$\\
0001 & $\begin{cases}
\left|\Psi_{0}\right\rangle \otimes\left|\Psi_{Id}\right\rangle \otimes\left|\Psi_{0}\right\rangle \\
\left|\Psi_0\right\rangle \otimes\left|\Psi_{not}\right\rangle \otimes\left|\Psi_{Id}\right\rangle \\
\left|\Psi_{1}\right\rangle \otimes\left|\Psi_{Id}\right\rangle \otimes\left|\Psi_{not}\right\rangle \\
\left|\Psi_{1}\right\rangle \otimes\left|\Psi_{not}\right\rangle \otimes\left|\Psi_{1}\right\rangle 
\end{cases}$\\  
0010 & $\begin{cases}
\left|\Psi_0\right\rangle \otimes\left|\Psi_{not}\right\rangle \otimes\left|\Psi_0\right\rangle \\
\left|\Psi_0\right\rangle \otimes\left|\Psi_{Id}\right\rangle \otimes\left|\Psi_{Id}\right\rangle \\
\left|\Psi_{1}\right\rangle \otimes\left|\Psi_{not}\right\rangle \otimes\left|\Psi_{not}\right\rangle \\
\left|\Psi_{1}\right\rangle \otimes\left|\Psi_{Id}\right\rangle \otimes\left|\Psi_{1}\right\rangle 
\end{cases}$ \\
0011 & $\begin{cases}
\left|\Psi_{0}\right\rangle \otimes\left|\Psi_{1}\right\rangle \otimes\left|\Psi_0\right\rangle \\
\left|\Psi_0\right\rangle \otimes\left|\Psi_0\right\rangle \otimes\left|\Psi_{Id}\right\rangle \\
\left|\Psi_{1}\right\rangle \otimes\left|\Psi_{1}\right\rangle \otimes\left|\Psi_{not}\right\rangle \\
\left|\Psi_{1}\right\rangle \otimes\left|\Psi_0\right\rangle \otimes\left|\Psi_{1}\right\rangle 
\end{cases}$ \\ 
0100 & $\begin{cases}
\left|\Psi_{Id}\right\rangle \otimes\left|\Psi_0\right\rangle \otimes\left|\Psi_0\right\rangle \\
\left|\Psi_{Id}\right\rangle \otimes\left|\Psi_{1}\right\rangle \otimes\left|\Psi_{Id}\right\rangle \\
\left|\Psi_{not}\right\rangle \otimes\left|\Psi_0\right\rangle \otimes\left|\Psi_{not}\right\rangle \\
\left|\Psi_{not}\right\rangle \otimes\left|\Psi_{1}\right\rangle \otimes\left|\Psi_{1}\right\rangle 
\end{cases}$ \\
0101 & $\begin{cases}
\left|\Psi_{Id}\right\rangle \otimes\left|\Psi_{Id}\right\rangle \otimes\left|\Psi_0\right\rangle \\
\left|\Psi_{Id}\right\rangle \otimes\left|\Psi_{not}\right\rangle \otimes\left|\Psi_{Id}\right\rangle \\
\left|\Psi_{not}\right\rangle \otimes\left|\Psi_{Id}\right\rangle \otimes\left|\Psi_{not}\right\rangle \\
\left|\Psi_{not}\right\rangle \otimes\left|\Psi_{not}\right\rangle \otimes\left|\Psi_{1}\right\rangle 
\end{cases}$\\ 
0110 & $\begin{cases}
\left|\Psi_{Id}\right\rangle \otimes\left|\Psi_{not}\right\rangle \otimes\left|\Psi_0\right\rangle \\
\left|\Psi_{Id}\right\rangle \otimes\left|\Psi_{Id}\right\rangle \otimes\left|\Psi_{Id}\right\rangle \\
\left|\Psi_{not}\right\rangle \otimes\left|\Psi_{not}\right\rangle \otimes\left|\Psi_{not}\right\rangle \\
\left|\Psi_{not}\right\rangle \otimes\left|\Psi_{Id}\right\rangle \otimes\left|\Psi_{1}\right\rangle 
\end{cases}$ \\
0111 & $\begin{cases}
\left|\Psi_{Id}\right\rangle \otimes\left|\Psi_{1}\right\rangle \otimes\left|\Psi_0\right\rangle \\
\left|\Psi_{Id}\right\rangle \otimes\left|\Psi_0\right\rangle \otimes\left|\Psi_{Id}\right\rangle \\
\left|\Psi_{not}\right\rangle \otimes\left|\Psi_{1}\right\rangle \otimes\left|\Psi_{not}\right\rangle \\
\left|\Psi_{not}\right\rangle \otimes\left|\Psi_0\right\rangle \otimes\left|\Psi_{1}\right\rangle 
\end{cases}$ \\ \hline
\end{tabular}
}%
\subfloat[]{%
\begin{tabular}{@{}c@{}@{}c@{}}
\hline 
Gate & Encoding states \\
\hline 
1000 & $\begin{cases}
\left|\Psi_{not}\right\rangle \otimes\left|\Psi_0\right\rangle \otimes\left|\Psi_0\right\rangle \\
\left|\Psi_{not}\right\rangle \otimes\left|\Psi_{1}\right\rangle \otimes\left|\Psi_{Id}\right\rangle \\
\left|\Psi_{Id}\right\rangle \otimes\left|\Psi_0\right\rangle \otimes\left|\Psi_{not}\right\rangle \\
\left|\Psi_{Id}\right\rangle \otimes\left|\Psi_{1}\right\rangle \otimes\left|\Psi_{1}\right\rangle 
\end{cases}$ \\
1001 & $\begin{cases}
\left|\Psi_{not}\right\rangle \otimes\left|\Psi_{Id}\right\rangle \otimes\left|\Psi_0\right\rangle \\
\left|\Psi_{not}\right\rangle \otimes\left|\Psi_{not}\right\rangle \otimes\left|\Psi_{Id}\right\rangle \\
\left|\Psi_{Id}\right\rangle \otimes\left|\Psi_{Id}\right\rangle \otimes\left|\Psi_{not}\right\rangle \\
\left|\Psi_{Id}\right\rangle \otimes\left|\Psi_{not}\right\rangle \otimes\left|\Psi_{1}\right\rangle 
\end{cases}$ \\
1010 & $\begin{cases}
\left|\Psi_{not}\right\rangle \otimes\left|\Psi_{not}\right\rangle \otimes\left|\Psi_0\right\rangle \\
\left|\Psi_{not}\right\rangle \otimes\left|\Psi_{Id}\right\rangle \otimes\left|\Psi_{Id}\right\rangle \\
\left|\Psi_{Id}\right\rangle \otimes\left|\Psi_{not}\right\rangle \otimes\left|\Psi_{not}\right\rangle \\
\left|\Psi_{Id}\right\rangle \otimes\left|\Psi_{Id}\right\rangle \otimes\left|\Psi_{1}\right\rangle 
\end{cases}$ \\
1011 & $\begin{cases}
\left|\Psi_{not}\right\rangle \otimes\left|\Psi_{1}\right\rangle \otimes\left|\Psi_0\right\rangle \\
\left|\Psi_{not}\right\rangle \otimes\left|\Psi_0\right\rangle \otimes\left|\Psi_{Id}\right\rangle \\
\left|\Psi_{Id}\right\rangle \otimes\left|\Psi_{1}\right\rangle \otimes\left|\Psi_{not}\right\rangle \\
\left|\Psi_{Id}\right\rangle \otimes\left|\Psi_0\right\rangle \otimes\left|\Psi_{1}\right\rangle 
\end{cases}$ \\
1100 & $\begin{cases}
\left|\Psi_{1}\right\rangle \otimes\left|\Psi_0\right\rangle \otimes\left|\Psi_0\right\rangle \\
\left|\Psi_{1}\right\rangle \otimes\left|\Psi_{1}\right\rangle \otimes\left|\Psi_{Id}\right\rangle \\
\left|\Psi_0\right\rangle \otimes\left|\Psi_0\right\rangle \otimes\left|\Psi_{not}\right\rangle \\
\left|\Psi_0\right\rangle \otimes\left|\Psi_{1}\right\rangle \otimes\left|\Psi_{1}\right\rangle 
\end{cases}$ \\
1101 & $\begin{cases}
\left|\Psi_{1}\right\rangle \otimes\left|\Psi_{Id}\right\rangle \otimes\left|\Psi_0\right\rangle \\
\left|\Psi_{1}\right\rangle \otimes\left|\Psi_{not}\right\rangle \otimes\left|\Psi_{Id}\right\rangle \\
\left|\Psi_0\right\rangle \otimes\left|\Psi_{Id}\right\rangle \otimes\left|\Psi_{not}\right\rangle \\
\left|\Psi_0\right\rangle \otimes\left|\Psi_{not}\right\rangle \otimes\left|\Psi_{1}\right\rangle 
\end{cases}$\\
1110 & $\begin{cases}
\left|\Psi_{1}\right\rangle \otimes\left|\Psi_{not}\right\rangle \otimes\left|\Psi_0\right\rangle \\
\left|\Psi_{1}\right\rangle \otimes\left|\Psi_{Id}\right\rangle \otimes\left|\Psi_{Id}\right\rangle \\
\left|\Psi_0\right\rangle \otimes\left|\Psi_{not}\right\rangle \otimes\left|\Psi_{not}\right\rangle \\
\left|\Psi_0\right\rangle \otimes\left|\Psi_{Id}\right\rangle \otimes\left|\Psi_{1}\right\rangle 
\end{cases}$\\
1111 & $\begin{cases}
\left|\Psi_{1}\right\rangle \otimes\left|\Psi_{1}\right\rangle \otimes\left|\Psi_0\right\rangle \\
\left|\Psi_{1}\right\rangle \otimes\left|\Psi_0\right\rangle \otimes\left|\Psi_{Id}\right\rangle \\
\left|\Psi_0\right\rangle \otimes\left|\Psi_{1}\right\rangle \otimes\left|\Psi_{not}\right\rangle \\
\left|\Psi_0\right\rangle \otimes\left|\Psi_0\right\rangle \otimes\left|\Psi_{1}\right\rangle 
\end{cases}$\\
\hline 
  \end{tabular}}}
  \caption{Encoding scheme for gates, using the three-photon, linear scheme.}
  \label{states_ell}
\end{table}
\vspace{7cm}
\begin{table}[h]

\captionsetup[subfloat]{labelformat=empty}
\subfloat[]{%
\begin{tabular}{c c}
\hline 
Gate & Encoding states \\ \hline 
0000 & $\begin{cases}
\left|\Psi_0\right\rangle \otimes\left|\Psi_{0}^{e}\right\rangle \\
\left|\Psi_{1}\right\rangle \otimes\left|\Psi_{4}^{e}\right\rangle 
\end{cases}$ \\
0001 & $\begin{cases}
\left|\Psi_0\right\rangle \otimes\left|\Psi_{1}^{e}\right\rangle \\
\left|\Psi_{1}\right\rangle \otimes\left|\Psi_{5}^{e}\right\rangle 
\end{cases}$ \\
0010 & $\begin{cases}
\left|\Psi_0\right\rangle \otimes\left|\Psi_{2}^{e}\right\rangle \\
\left|\Psi_{1}\right\rangle \otimes\left|\Psi_{6}^{e}\right\rangle 
\end{cases}$  \\
0011 & $\begin{cases}
\left|\Psi_0\right\rangle \otimes\left|\Psi_{3}^{e}\right\rangle \\
\left|\Psi_{1}\right\rangle \otimes\left|\Psi_{7}^{e}\right\rangle 
\end{cases}$ \\
0100 & $\begin{cases}
\left|\Psi_{Id}\right\rangle \otimes\left|\Psi_{0}^{e}\right\rangle \\
\left|\Psi_{not}\right\rangle \otimes\left|\Psi_{4}^{e}\right\rangle 
\end{cases}$  \\
0101 & $\begin{cases}
\left|\Psi_{Id}\right\rangle \otimes\left|\Psi_{1}^{e}\right\rangle \\
\left|\Psi_{not}\right\rangle \otimes\left|\Psi_{5}^{e}\right\rangle 
\end{cases}$  \\
0110 & $\begin{cases}
\left|\Psi_{Id}\right\rangle \otimes\left|\Psi_{2}^{e}\right\rangle \\
\left|\Psi_{not}\right\rangle \otimes\left|\Psi_{6}^{e}\right\rangle 
\end{cases}$  \\
0111 & $\begin{cases}
\left|\Psi_{Id}\right\rangle \otimes\left|\Psi_{3}^{e}\right\rangle \\
\left|\Psi_{not}\right\rangle \otimes\left|\Psi_{7}^{e}\right\rangle 
\end{cases}$  \\
\hline
\end{tabular}
} 
\subfloat[]{%
\begin{tabular}{c c}
\hline 
Gate & Encoding states \\ \hline 
1000 & $\begin{cases}
\left|\Psi_{not}\right\rangle \otimes\left|\Psi_{0}^{e}\right\rangle \\
\left|\Psi_{Id}\right\rangle \otimes\left|\Psi_{4}^{e}\right\rangle 
\end{cases}$ \\
1001 & $\begin{cases}
\left|\Psi_{not}\right\rangle \otimes\left|\Psi_{1}^{e}\right\rangle \\
\left|\Psi_{Id}\right\rangle \otimes\left|\Psi_{5}^{e}\right\rangle 
\end{cases}$ \\
1010 & $\begin{cases}
\left|\Psi_{not}\right\rangle \otimes\left|\Psi_{2}^{e}\right\rangle \\
\left|\Psi_{Id}\right\rangle \otimes\left|\Psi_{6}^{e}\right\rangle 
\end{cases}$ \\
1011 & $\begin{cases}
\left|\Psi_{not}\right\rangle \otimes\left|\Psi_{3}^{e}\right\rangle \\
\left|\Psi_{Id}\right\rangle \otimes\left|\Psi_{7}^{e}\right\rangle 
\end{cases}$ \\
1100 & $\begin{cases}
\left|\Psi_{1}\right\rangle \otimes\left|\Psi_{0}^{e}\right\rangle \\
\left|\Psi_0\right\rangle \otimes\left|\Psi_{4}^{e}\right\rangle 
\end{cases}$ \\
1101 & $\begin{cases}
\left|\Psi_{1}\right\rangle \otimes\left|\Psi_{1}^{e}\right\rangle \\
\left|\Psi_0\right\rangle \otimes\left|\Psi_{5}^{e}\right\rangle 
\end{cases}$ \\
1110 & $\begin{cases}
\left|\Psi_{1}\right\rangle \otimes\left|\Psi_{2}^{e}\right\rangle \\
\left|\Psi_0\right\rangle \otimes\left|\Psi_{6}^{e}\right\rangle 
\end{cases}$ \\
1111 & $\begin{cases}
\left|\Psi_{1}\right\rangle \otimes\left|\Psi_{3}^{e}\right\rangle \\
\left|\Psi_0\right\rangle \otimes\left|\Psi_{7}^{e}\right\rangle 
\end{cases}$ \\ \hline
\end{tabular}}
\caption{Encoding scheme for gates, using the two-photon, elliptical scheme.}\label{states_lin}
\end{table}
\newpage
\begin{table}[t]

\centering
\begin{tabular}{l r}
\hline
\textbf{State} & \textbf{Fidelity}  \\  \hline 
$\ket{\Psi_0}$ \hspace{2em}	& 0.994	$\pm$ 0.006\\
$\ket{\Psi_{Id}}$ & 0.995	$\pm$ 0.002\\
$\ket{\Psi_{not}}$ & 0.997	$\pm$ 0.005\\
$\ket{\Psi_1}$ & 0.998	$\pm$ 0.003\\
$\ket{\Psi_{0}^{e}}$	& 0.996	$\pm$ 0.002\\
$\ket{\Psi_{1}^{e}}$& 0.997	$\pm$ 0.003\\
$\ket{\Psi_{2}^{e}}$& 0.992	$\pm$ 0.002\\
$\ket{\Psi_{3}^{e}}$& 0.997	$\pm$ 0.002\\
$\ket{\Psi_{4}^{e}}$& 0.993	$\pm$ 0.002\\
$\ket{\Psi_{5}^{e}}$	& 0.997	$\pm$ 0.001\\
$\ket{\Psi_{6}^{e}}$	& 0.991	$\pm$ 0.007\\
$\ket{\Psi_{7}^{e}}$& 0.991	$\pm$ 0.008\\ \hline
\end{tabular}
\caption{ Quantum state fidelity of all used single-qubit states. The error is estimated using a 500-cycle Monte-Carlo simulation with Poissonian noise added to the experimental counts.}
\label{Fidelities}
\end{table}


\putbib[OTP_bib]
\end{bibunit}

\end{document}